\documentclass[fleqn,usenatbib]{mnras}

\usepackage{newtxtext,newtxmath}

\usepackage[T1]{fontenc}

\DeclareRobustCommand{\VAN}[3]{#2}
\let\VANthebibliography\thebibliography
\def\thebibliography{\DeclareRobustCommand{\VAN}[3]{##3}\VANthebibliography}

\usepackage{graphicx}	
\usepackage{amsmath}	
\usepackage{float}

\title[The damping of MHD turbulence]{Damping of MHD Turbulence in A Partially Ionized Medium}

\author[Hu et al.]{Yue Hu$^{1,2}$\thanks{E-mail: yue.hu@wisc.edu; }, Siyao Xu$^{3}$\thanks{E-mail: sxu@ias.edu (NASA Hubble Fellow)}, Lev Arzamasskiy$^{3}$, James M. Stone$^{3,4}$, A. Lazarian $^{2,5}$
\\
$^{1}$Department of Physics, University of Wisconsin-Madison, Madison, WI 53706, USA\\
$^{2}$Department of Astronomy, University of Wisconsin-Madison, Madison, WI 53706, USA\\
$^{3}$Institute for Advanced Study, 1 Einstein Drive, Princeton, NJ 08540, USA\\
$^{4}$Department of Astrophysical Sciences, Princeton University, Princeton, NJ 08544, USA\\
$^{5}$ Centro de Investigación en Astronomía, Universidad Bernardo O’Higgins, Santiago, General Gana 1760, 8370993, Chile\\
}

\date{Accepted XXX. Received YYY; in original form ZZZ}

\pubyear{2023}

\begin{document}
\label{firstpage}
\pagerange{\pageref{firstpage}--\pageref{lastpage}}
\maketitle

\begin{abstract}
The coupling state between ions and neutrals in the interstellar medium plays a key role in the dynamics of magnetohydrodynamic (MHD) turbulence, but is challenging to study numerically. In this work, we investigate the damping of MHD turbulence in a partially ionized medium using 3D two-fluid (ions+neutrals) simulations generated with the AthenaK code. Specifically, we examine the velocity, density, and magnetic field statistics of the two-fluid MHD turbulence in different regimes of neutral-ion coupling. Our results demonstrate that when ions and neutrals are strongly coupled, the velocity statistics resemble those of single-fluid MHD turbulence. Both the velocity structures and kinetic energy spectra of ions and neutrals are similar,  while their density structures can be significantly different. With an excess of small-scale sharp density fluctuations in ions, the density spectrum in ions is shallower than that of neutrals. When ions and neutrals are weakly coupled, the turbulence in ions is more severely damped due to the ion-neutral collisional friction than that in neutrals, resulting in a steep kinetic energy spectrum and density spectrum in ions compared to the Kolmogorov spectrum. We also find that the magnetic energy spectrum basically follows the shape of the kinetic energy spectrum of ions, irrespective of the coupling regime. In addition, we find large density fluctuations in ions and neutrals and thus spatially inhomogeneous ionization fractions. As a result, the neutral-ion decoupling and damping of MHD turbulence take place over a range of length scales. 
\end{abstract}




\section{Introduction}
Magnetohydrodynamic (MHD) turbulence is an essential element in the multi-phase interstellar medium (ISM; \citealt{1981MNRAS.194..809L,1995ApJ...443..209A,2004ARAA..42..211E,MO07,2009ApJ...693..250B,2010ApJ...710..853C,2019ApJ...872...10X,2022ApJ...934....7H}). To date, extensive numerical studies have focused on the properties of MHD turbulence in a fully-ionized single-fluid regime \citep{2002ApJ...564..291C,2003MNRAS.345..325C,2007ApJ...658..423K,2013ApJ...771..123B,2013MNRAS.436.1245F,2016ApJ...825..154Z,2016PhRvL.117w5101K,HXL21,2021NatAs...5..342M,2022MNRAS.513.2100H}. However, the multiphase ISM has a wide range of ionization fractions \citep{1978ppim.book.....S,1989ApJ...345..782M, 2011piim.book.....D,2014MNRAS.439.2197M, 2021ApJ...912....7P}, and thus MHD turbulence should be considered in two fluids, i.e., ionized and neutral fluids. In a weakly ionized medium, at large scales, the single-fluid treatment is valid when ions and neutrals are strongly coupled via their frequent collisions. At scales smaller than the neutral-ion decoupling scale, the collisional coupling of ions and neutrals becomes weak and neutrals start to decouple from ions. The coupling state between ions and neutrals can significantly affect the dynamics of MHD turbulence and result in damping of its energy cascade \citep{1965RvPP....1..205B,1978ApJ...225...95L,1983ApJ...270..511Z,1996ApJ...465..775B}. 

MHD turbulence in partially interstellar phases regulates many key astrophysical processes and phenomena, such as star formation \citep{1956MNRAS.116..503M,1972ApJ...173...87N,1976ApJ...210..326M,1979ApJ...228..159M,1989ApJ...342..834L, MK04,MO07,2010ApJ...720.1612M,2012ApJ...761..156F,2020ApJ...899..115X,2022MNRAS.513.2100H}, linewidth difference between ions and neutrals \citep{2008ApJ...677.1151L,2010ApJ...718..905L,2015ApJ...810...44X}, density filament formation \citep{2019ApJ...878..157X}, cosmic ray propagation \citep{2016ApJ...826..166X,2021ApJ...914....3P,2022ApJ...927...94X,2023MNRAS.519.1503S}, turbulent dynamo \citep{2016ApJ...833..215X,2019ApJ...872...62X,2019PhRvF...4b4608B}, injection of turbulence in very local ISM \citep{2022ApJ...941L..19X}, heating of solar chromosphere \citep{2016ApJ...819L..11S}. In view of this importance, there has been significant effort in studying MHD turbulence in the presence of neutrals. Earlier analytical studies were mainly focused on linear MHD waves in a weakly ionized medium and their ion-neutral collisional damping \citep{1969ApJ...156..445K,1988ApJ...332..984F,1996ApJ...465..775B,2004A&A...422.1073K,2007A&A...461..731F,2011A&A...529A..82Z,2011arXiv1103.6037M}. However, unlike MHD waves, MHD turbulence is highly non-linear and dynamical \citep{GS95,LV99}. In a compressible medium, it consists of energy cascades of three fundamental modes (Alfv\'en, fast, and slow), rather than a collection of linear MHD waves \citep{2003MNRAS.345..325C}. The effects of ion-neutral collisions and the resulting damping of the cascade of compressible MHD turbulence were analytically investigated in \cite{2001ApJ...562..279L,2004ApJ...603..180L,2015ApJ...810...44X,2016ApJ...826..166X,2018arXiv181008205X}. In addition to the local physical conditions, the properties of MHD turbulence and turbulence parameters are important for determining the damping effect. With the recent development in theories, simulations, and observations of MHD turbulence \citep{2019tuma.book.....B}, our understanding of the dynamics and scaling properties of MHD turbulence has been significantly improved. Along the energy cascade of MHD turbulence, its anisotropy increases with decreasing length scales \citep{GS95,LV99,2000ApJ...539..273C,2002ApJ...564..291C}. The ambipolar diffusion scale derived using the wave description of MHD turbulence or isotropic turbulence scaling cannot provide a proper estimate of the damping scale of MHD turbulence \citep{2015ApJ...810...44X}. The actual ambipolar diffusion scale, i.e., ion-neutral collisional damping scale, can be smaller due to the turbulence. In addition, unlike infinitesimal perturbations around an equilibrium state for MHD waves, super-Aflv\'enic and super-sonic turbulence in neutral-dominated cold interstellar phases can induce magnetic and density fluctuations much larger than their mean values \citep{2012ApJ...761..156F,2022arXiv221011023H}. The magnetic field and density inhomogeneity significantly complicate the analysis of the damping of MHD turbulence.

Simulating two-fluid MHD turbulence is more challenging than single-fluid MHD simulations due to the high Alfv\'en speed of ions at low ionization fractions, which requires a much smaller time step. To address this issue, the "heavy-ion approximation" (HIA) has been adopted to accelerate explicit two-fluid MHD simulations \citep{2006ApJ...638..281O,2006ApJ...653.1280L,2010ApJ...720.1612M}. This approach increases the mass of ions and reduces the ion-neutral drag coefficient $\gamma_{\rm d}$ \citep{1983ApJ...264..485D,1992pavi.book.....S} accordingly. However, for simulating MHD turbulence in a weakly ionized medium, the HIA approximation may raise uncertainties \citep{2018SSRv..214...58B,2010MNRAS.406.1201T}. The single-fluid treatment used in, e.g., \cite{2006MNRAS.366.1329O,2007MNRAS.376.1648O} for numerical modeling of MHD turbulence in a weakly ionized medium cannot fully capture the two-fluid effects in the weakly coupled regime \citep{2010MNRAS.406.1201T,2016ApJ...826..166X}. Despite these challenges, numerical methods are crucial for testing theories of two-fluid MHD turbulence and studying ion-neutral collisional damping in an inhomogeneous medium. Three-dimensional (3D) simulations of two-fluid MHD turbulence with the RIEMANN code \citep{1998ApJS..116..119B} have been carried out by \cite{2010MNRAS.406.1201T,2014MNRAS.439.2197M}. These studies show differences in the turbulent energy spectra of ions and neutrals. The persistence of the energy cascade of Alfv\'en modes on scales smaller than the amplipolar diffusion scale calculated using the wave description of MHD turbulence \citep{2015ApJ...805..118B} suggests that the damping of MHD turbulence is different from the damping of MHD waves.

In this work, we use 3D two-fluid MHD simulations to test the theoretical models developed by \cite{2015ApJ...810...44X,2016ApJ...826..166X} and study the properties of MHD turbulence in various ion-neutral coupling regimes in the presence of turbulence-induced density inhomogeneities.  We perform the two-fluid MHD turbulence simulations using MHD code Athena++ \citep{2020ApJS..249....4S}, updated using the Kokkos framework (denoted as AthenaK). The code utilizes IMEX integrators \citep{PareschiRusso2005} to enable high-order (in time) implementation of ion-neutral drag terms, which allows for higher accuracy and stability than operator-split methods ({\color{blue}Arzamasskiy \& Stone, in prep.}). To reduce the computational cost, we consider a moderately low ionization fraction. Different regimes of ion-neutral coupling are achieved by varying the numerical value of $\gamma_{\rm d}$. To evaluate the limitations of this approach, we also carry out simulations with the same $\gamma_{\rm d} \rho_i$ but a lower ionization fraction, where $\rho_i$ is the ion mass density. 


The paper is organized as follows: in \S~\ref{sec:data}, we describe the 3D numerical simulations of two-fluid MHD turbulence used in this study. In \S~\ref{sec:theory}, we review the recent theoretical understanding on neutral-ion decoupling and collisional damping of MHD turbulence. In \S~\ref{sec:result}, we present the numerical results on the statistics of the velocity, density, and magnetic field in both ions and neutrals. The implications of the results and comparison with earlier studies are discussed in  \S~\ref{sec:dis}, and our main findings are summarized in \S~\ref{sec:con}.

\section{Numerical simulation}
\label{sec:data}
\subsection{Numerical setup}
The 3D two-fluid simulations analyzed in this work are generated using the Athena++ implemented with Kokkos \citep{2020ApJS..249....4S}. We consider the two-fluid magneto-fluid system, comprised of ions (together with electrons) and neutrals. The effects of gravity, heat conduction, ionization, and recombination are not included in the current study. The simulations solve the ideal two-fluid MHD equations, using periodic boundary conditions, IMEX3 time integration algorithm, and PPM4 spatial reconstruction method. The equations are:
\begin{equation}
\label{eq.mhd}
\begin{aligned}
    &\partial\rho_i/\partial t +\nabla\cdot\left(\rho_i\pmb{v}_i\right)=0,\\
    \\
    &\partial\rho_n/\partial t +\nabla\cdot(\rho_n\pmb{v}_n)=0,\\ 
    \\
    &\partial(\rho_i\pmb{v}_i)/\partial 
t+\nabla\cdot\left[\rho_i\pmb{v}_i\pmb{v}_i^T+\left(c_s^2\rho_i+\frac{B^2}{8\pi}\right)\pmb{I}-\frac{\pmb{B}\pmb{B}^T}{4\pi}\right ]\\
    &~~~~~~~~~~~~~~~~~~~~~~~~~~~~~~~~~~~~~=\gamma_{\rm d}\rho_n\rho_i(\pmb{v}_n-\pmb{v}_i)+\pmb{f}_i,\\     
    \\
    &\partial(\rho_n\pmb{v}_n)/\partial t+\nabla\cdot\left[\rho_n\pmb{v}_n\pmb{v}_n^T+c_s^2\rho_n\pmb{I}\right]
    =\gamma_{\rm d}\rho_n\rho_i(\pmb{v}_i-\pmb{v}_n)+\pmb{f}_n,\\
    \\
&\partial\pmb{B}/\partial t-\nabla\times(\pmb{v}_i\times\pmb{B})=0,\\
    \\
    &\nabla \cdot\pmb{B}=0,\\
    \end{aligned}
\end{equation}
here $\rho$ and $v$ are the mass density and velocity of the ionized fluid (with the subscript "$i$") and neutral fluid (with the subscript "$n$"), respectively. We adopt an isothermal equation of state, where the sound speed $c_s$ is constant. The isothermal condition applies to a medium with efficient cooling, such as molecular clouds \citep{2010MNRAS.406.1201T,2014MNRAS.439.2197M}. It only breaks down when density exceeds $\sim 10^{10} {\rm cm}^{-3}$ \citep{2012ApJ...758...86F}. The ion-neutral collisional damping under other conditions will be studied in our future work.To drive turbulent motions in ions and neutrals, a stochastic forcing term $\pmb{f}$ is utilized. Explicitly, $\pmb{f}_i$ and $\pmb{f}_n$ are weighted by ion and neutral densities to achieve the same injected turbulent velocities in the two fluids.

At the start of the simulation, the magnetic field and (ion and neutral) density fields are set to be uniform, with the magnetic field along the $z$-axis. The initial ionization fraction is $\xi_i=\rho_i/(\rho_i+\rho_n)$, where $\rho_i$ and $\rho_n$ are the initial mass densities of ions and neutrals. The simulation box is divided into 480$^3$ cells and is uniformly staggered.

\begin{table}
	\centering
	\label{tab.1}
	\begin{tabular}{| c | c | c | c | c | c | c | c |}
		\hline
		Run & $M_s$ & $M_A$ & $\beta$ & $\gamma_{\rm d}$ & $\xi_i$ & $k_{\rm dec, \parallel}$ & $k_{\rm dec, \bot}$\\ \hline\hline
		$\gamma5$ & 1.10 & 1.07 & 1.9  & 5 & 0.1 & 0.5 & 0.3 \\
		$\gamma25$ & 1.06 & 1.08 & 2.1 & 25 & 0.1 & 3 & 4\\
		$\gamma100$ & 0.95 & 0.97 & 2.1 & 100 & 0.1 & 10 & 32 \\
		$\gamma250$ & 1.13 & 1.12 & 2.0  & 250 & 0.01 & 3 & 4 \\
		$\gamma1e3$ & 1.19 & 1.07 & 1.6 & $10^3$ &0.1 & $10^2$ & $10^3$ \\
		$\gamma1e4$ & 0.97 & 0.87 & 1.6 & $10^4$ &0.1 & $10^3$ & 3$\times10^4$ \\
		$\gamma1e5$ & 1.05 & 0.91 & 1.5 & $10^5$ & 0.1 & $10^4$ & $10^6$ \\
		\hline
	\end{tabular}
	\caption{Setups of two-fluid simulations. $M_s$ and $M_A$ are the instantaneous RMS values at each snapshot that is taken. $\beta=2(M_A/M_s)^2$ is plasma compressibility. $k_{\rm dec, \parallel}$ and $k_{\rm dec, \bot}$ are theoretically expected parallel and perpendicular components of the neutral-ion decoupling wavenumber, respectively. The listed $\gamma_{\rm d}$ is given in numerical units. To obtain a dimensionless value, divide $\gamma_{\rm d}$ by $v_{\rm inj}/(L_{\rm inj}\rho_i)$, which is fixed at 10 and 100 (in numerical units) for $\xi_i=0.1$ and $0.01$, respectively.}
\end{table}

\subsection{Turbulence driving}
The forcing term, $\pmb{f}$, is introduced to drive the turbulence in a solenoidal manner. This is ensured by making the forcing term divergence-free. The forcing term is modeled using the stochastic Ornstein-Uhlenbeck (OU) process, which allows us to control the auto-correlation timescale, $t_{\rm c}$, of the turbulence. The auto-correlation timescale is approximately equal to $t_{\rm c}\approx L_{\rm inj}/(2\pi v_{\rm A})$, where $L_{\rm inj}$ is the turbulence injection scale and $v_{A} = \frac{B}{\sqrt{4\pi(\rho_i + \rho_n)}}$ is the Alfv\'en speed in the two fluids. The time step is the minimum time step allowed by the Courant–Friedrichs–Lewy stability condition for the ion and neutral fluids, respectively.

To vary the level of turbulence, we change the values of $v_{\rm inj}$. The energy injection is focused around wavenumber $k=2\pi/l=1-2$ (in the unit of $2\pi/L_{\rm box}$, where $L_{\rm box}$ is the length of simulation box) in Fourier space, where $l$ is the length scale in real space. The turbulence is numerically dissipated at wavenumber $k_{\rm dis}\approx40-50$. We run the simulations for six eddy turnover times to ensure that the turbulence has reached a statistically stable state.


The simulation of scale-free turbulence can be characterized by the sonic Mach number, $M_s = \frac{v_{\rm inj}}{c_{s}}$, and the Alfv\'en Mach number, $M_A = \frac{v_{\rm inj}}{v_{A}}$, where $v_{\rm inj}$ is the injection velocity. In this work, we fix $M_s$ and $M_A$ to approximate unity, ensuring that the simulations fully fall into the strong turbulence regime \footnote{The strong turbulence regime has the "critical balance" condition \citep{GS95} satisfied. See \S~\ref{sec:theory}., where the magnetic field becomes dynamically important and turbulence anisotropy develops \citep{2006ApJ...645L..25L}. This is defined in the range as [$l_{\rm dis}$, $L_{\rm inj}M_A^{-3}$] for super-Alfv\'enic ($M_A > 1$) and [$l_{\rm dis}, L_{\rm inj}M_A^{2}$] for sub-Alfv\'enic ($M_A < 1$).}  Here $l_{\rm dis}$ is the turbulence dissipation scale. The critical parameters for this study are listed in Tab.~{\color{blue} 1}.


\section{Theoretical Consideration}
\label{sec:theory}
\subsection{Anisotropic MHD turbulence}
\label{subsec:mhd}
Our understanding of MHD turbulence has undergone significant changes over the past few decades. MHD turbulence was initially considered to be isotropic despite the existence of magnetic fields \citep{1963AZh....40..742I,1965PhFl....8.1385K}. However, numerous numerical studies \citep{1981PhFl...24..825M,1983JPlPh..29..525S,1984ApJ...285..109H,1965PhFl....8.1385K,1995ApJ...447..706M,2000ApJ...539..273C,2001ApJ...554.1175M,2002ApJ...564..291C,2010ApJ...720..742K,HLX21a} and in situ measurements of solar wind \citep{2016ApJ...816...15W,2021ApJ...915L...8D,2020FrASS...7...83M} have revealed that the turbulence is anisotropic, rather than isotropic, when the effect of magnetic fields is non-negligible.

Fundamental work on anisotropic incompressible MHD turbulence theory was initiated by \citet{GS95} in the trans-Alfv\'enic regime with $M_A \approx 1$. \citet{GS95} found the "critical balance" condition, which equates the turbulence cascading time ($k_\bot v_l$)$^{-1}$ with the Alfv\'en wave period ($k_\parallel v_A$)$^{-1}$. Here, $k_\bot$ is the wavevector perpendicular to the magnetic field, and $v_l$ is the turbulent velocity at scale $l$.  Later studies further found that the "critical balance" condition can be valid in the strong turbulence regime when $M_A$ is not unity \citep{LV99,2006ApJ...645L..25L}: (i) in super-Alfv\'enic turbulence, where $M_A > 1$, the magnetic field's role at the injection scale, $L_{\rm inj}$, is insignificant, resulting in isotropic turbulence. However, as the turbulence cascades to smaller scales with decreasing turbulent velocity, the Alfv\'en speed becomes comparable to the turbulent speed at the Alfv\'en scale, $l_A=L_{\rm inj}M_A^{-3}$. This leads to the development of strong turbulence on smaller scales. (ii) in sub-Alfv\'enic turbulence ($M_A < 1$), the strong turbulence regime spans from the transitional scale $l_{\rm trans}=L_{\rm inj}M_A^{2}$ to smaller scales. Turbulence within the range from $L_{\rm inj}$ to $l_{\rm trans}$ is termed weak Alfv\'enic turbulence, which is wave-like and does not obey the "critical balance".

As turbulence cascades preferentially along the direction perpendicular to the local magnetic field \citep{LV99}, where the resistance to turbulent mixing of magnetic fields is the minimum, we have the scaling relations of velocity fluctuation in the strong turbulence regime:
\begin{equation}
\label{eq.lv99}
\begin{aligned}
        v_l &=
        \begin{cases}
        (\frac{l_\bot}{L_{\rm inj}})^{1/3}v_{\rm inj},~~~M_A>1\\
        (\frac{l_\bot}{L_{\rm inj}})^{1/3}v_{\rm inj}M_A^{1/3},~~~M_A<1\\
        \end{cases},
\end{aligned}
\end{equation}
where $l_\bot$ is the length scale perpendicular to the local magnetic field. By using the scaling of velocity fluctuation and the "critical balance" condition, one can easily obtain the anisotropy scaling:
\begin{equation}
\label{eq.gs95}
\begin{aligned}
        k_\parallel &=
        \begin{cases}
        (k_\bot L_{\rm inj})^{2/3}L_{\rm inj}^{-1}M_A,~~~M_A>1\\
        (k_\bot L_{\rm inj})^{2/3}L_{\rm inj}^{-1}M_A^{4/3},~~~M_A<1\\
        \end{cases}.
\end{aligned}
\end{equation}
 The scale-dependent anisotropy of MHD turbulence described by the above expression indicates that the turbulent eddy has a parallel size much larger than its perpendicular size, and this anisotropy increases with the decrease of length scales. However, note that the scale-dependent anisotropy can only be measured in the reference frame of local magnetic fields that percolate the turbulent eddy \citep{LV99,2000ApJ...539..273C}.

\begin{figure*}
\includegraphics[width=0.99\linewidth]{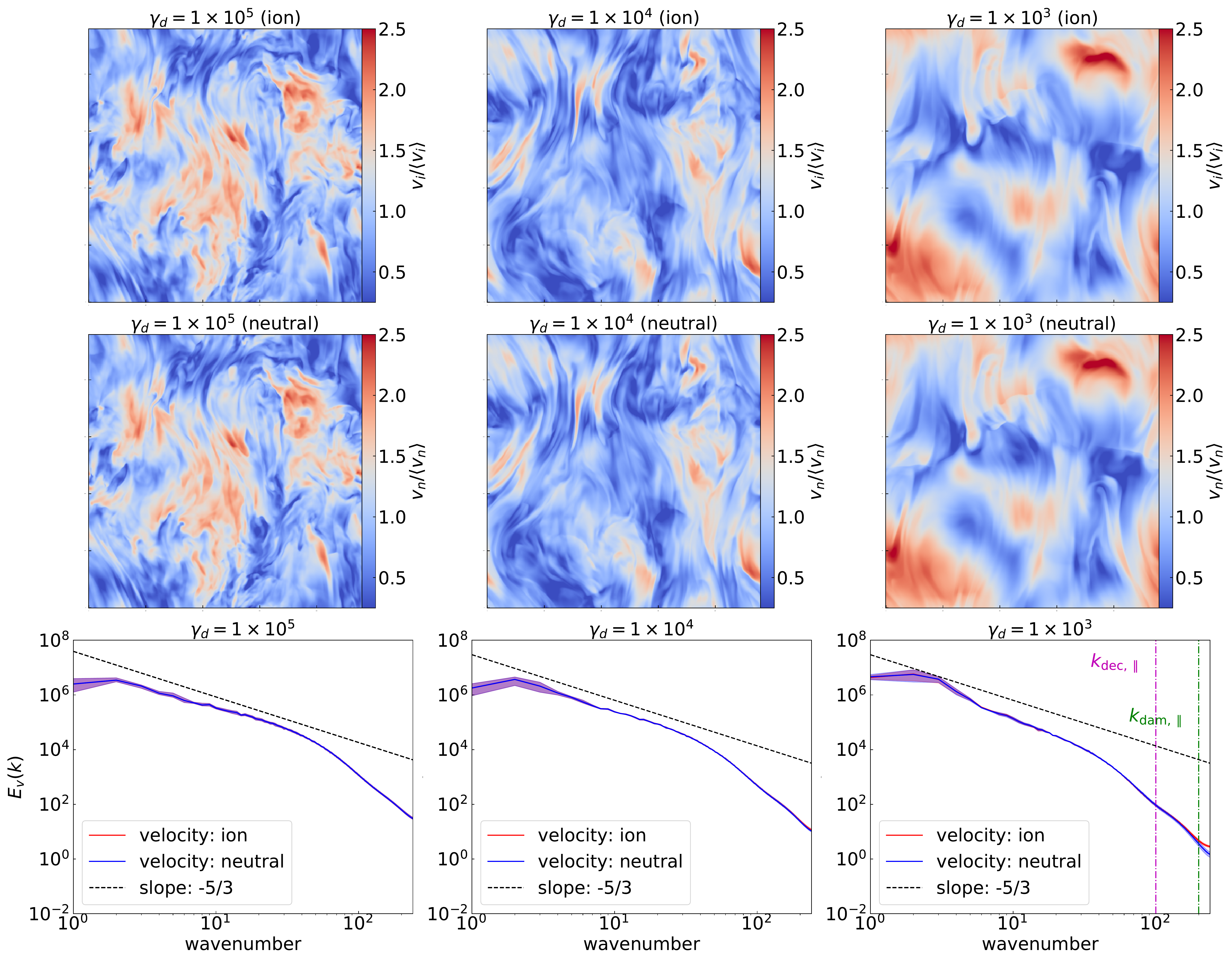}
    \caption{\textbf{Top and middle:} 2D slices (taken at $x=240$ cell) of ions' (top) and neutrals' (middle) velocity field. The velocity maps are normalized by the mean value. The view direction is perpendicular to the mean magnetic field, which is along the vertical $z$-direction. \textbf{Bottom:} Turbulent kinetic energy spectra of ions (red) and neutrals (blue). The spectra are averaged over several snapshots after turbulence reaches a statistically steady state, with the time interval equal to the largest eddy turnover time. The shadowed areas represent the variations. Magenta and green lines represent the theoretically expected neutral-ion parallel decoupling (Eq.~\ref{eq:decouple}) and damping wavenumbers (Eq.~\ref{eq.kc}), respectively, for Alfv\'en modes of MHD turbulence. The perpendicular decoupling and damping wavenumbers are not shown because they are larger than $10^3$.}
    \label{fig:v_map1}
\end{figure*}

\begin{figure*}
\includegraphics[width=0.99\linewidth]{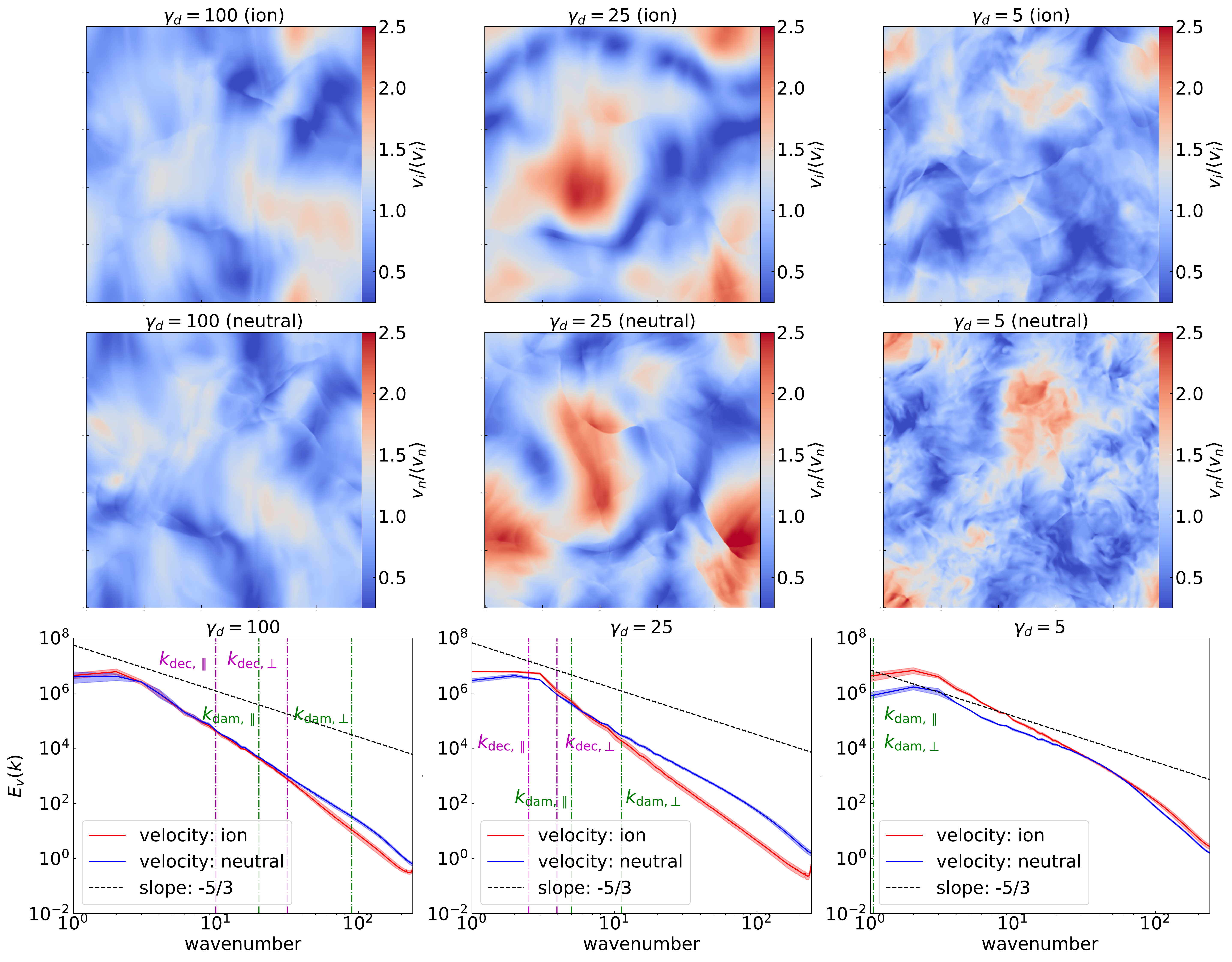}
    \caption{Same as Fig.~\ref{fig:v_map1}, but for $\gamma_{\rm d}=100,25,5$. }
    \label{fig:v_map2}
\end{figure*}

\subsection{Decoupling of ions and neutrals}
The interaction between ions and neutrals can be quantified by the neutral-ion collisional frequency $\nu_{ni} = \gamma_{\rm d} \rho_i=\gamma_{\rm d}\xi_i(\rho_i+\rho_n)$ and ion-neutral collisional frequency $\nu_{in} = \gamma_{\rm d} \rho_n$, respectively \citep{1992pavi.book.....S}. Neutrals start to decouple from ions when the energy cascading rate of MHD turbulence matches $\nu_{ni}$. Given $\nu_{in} \gg \nu_{ni}$ in a weakly ionized medium, ions decouple from neutrals on a much smaller scale than the neutral-ion decoupling scale, so here we mainly consider neutral-ion decoupling. The coupling status between ions and neutrals in MHD turbulence can be separated into three important regimes: (i) strongly coupled regime, in which the scales are larger than the neutral-ion decoupling scale. Neutrals and ions act as single-fluid in this regime. (ii) weakly coupled regime, where the scales are smaller than the
neutral-ion decoupling scale but larger than the ion-neutral decoupling scale.
Neutrals thus decouple from ions, but ions still couple to neutrals. (iii) Decoupled regime, in which the scales are smaller than the ion-neutral decoupling scale. Neutrals and ions in this regime are fully decoupled so if the turbulence injection happens in this regime, neutrals develop an independent hydrodynamic turbulent cascade and ions develop an MHD turbulent cascade. 

In earlier linear analysis \citep{1969ApJ...156..445K}, it was considered that the decoupling of neutrals from Alfv\'en wave oscillations at the neutral-ion decoupling wavenumber $k_{{\rm dec},\parallel}$. It can be determined by equating the Alfv\'en wave frequency and $\nu_{ni}$ \citep{1992pavi.book.....S}: 
\begin{equation}
k_{{\rm dec},\parallel} v_A=\nu_{ni},
\end{equation}
where the subscript "$\parallel$" means the wavevector parallel to the magnetic field. MHD turbulence was previously modeled as a collection of linear waves \citep{1999ApJ...520..204G}, and $k_{{\rm dec},\parallel} v_A=\nu_{ni}$ was taken as the neutral-ion decoupling wavenumber or ambipolar diffusion wavenumber of MHD turbulence \citep{1969ApJ...156..445K,1991ApJ...371..296M,2013A&A...560A..68H}. However, it is essential to note that MHD turbulence is a highly nonlinear phenomenon and the Alfv\'en wave-like motion in the strong turbulence regime with the critical balance cannot survive for more than a wave period. With the dynamically coupled turbulent mixing motion in the perpendicular direction and the wave-like motion in the parallel direction, scale-dependent anisotropy is one of its most important properties (see \S~\ref{subsec:mhd}).

For Alfv\'enic turbulence, which usually carries most of the MHD turbulence energy \citep{2002ApJ...564..291C,HLX21a}, the anisotropy suggests that the neutral-ion decoupling scale is not isotropic. The parallel component of the decoupling scale can be much larger than the perpendicular component when it is significantly smaller than $L_{\rm inj}$. Taking into account the critical-balance relation between turbulent motions and wave-like motions and the anisotropy of MHD turbulence, \cite{2015ApJ...810...44X} derived the parallel decoupling wavenumber $k_{{\rm dec},\parallel}$ and perpendicular decoupling wavenumber $k_{{\rm dec},\bot}$ by using the anisotropic scaling (Eq.~\ref{eq.gs95}) in strong MHD turbulence regimes (i.e., $k_{{\rm dec},\bot}>l_A^{-1}$ or $k_{{\rm dec},\bot}>l_{\rm trans}^{-1}$):
\begin{equation}
\label{eq:decouple}
\begin{aligned}
        k_{{\rm dec},\parallel}&=\nu_{\rm ni}v_A^{-1},\\
        k_{{\rm dec},\bot}&=
        \begin{cases}
        \nu_{\rm ni}^{3/2}L_{\rm inj}^{1/2}v_{\rm inj}^{-3/2},~~~M_A>1\\
        \nu_{\rm ni}^{3/2}L_{\rm inj}^{1/2}v_{\rm inj}^{-3/2}M_A^{-1/2},~~~M_A<1\\
        \end{cases}.
\end{aligned}
\end{equation}
Here, we only consider Alfv\'en turbulence as it carries most of the turbulent energy \citep{2003MNRAS.345..325C,2022MNRAS.tmp..341H}. For a more in-depth discussion on neutral-ion decoupling scales of the three MHD modes (Alfv\'en, fast, and slow), see \cite{2015ApJ...810...44X,2016ApJ...826..166X}.

\subsection{Neutral-ion collisional damping of MHD turbulence in a partially ionized medium}
At length scales larger than $k^{-1}_{{\rm dec},\parallel}$, ions and neutrals are perfectly coupled, and together carry the MHD turbulence. However, at length scales smaller than $k^{-1}_{{\rm dec},\bot}$, neutrals begin to decouple from ions, resulting in the development of their own hydrodynamic turbulent cascade, while ions continue to undergo frequent collisions with surrounding neutrals (down to the ion-neutral decoupling scale). As a result, the remaining MHD turbulence in ions is strongly affected and damped by the collisional friction exerted by neutrals, which is denoted as neutral-ion (collisional) damping.

When neutral-ion damping dominates over the damping caused by the kinematic viscosity of neutrals, the parallel damping wavenumber $k_{\rm dam,\parallel}$ for Alfv\'enic turbulence, as derived in \cite{2015ApJ...810...44X} by equating the turbulent cascading rate $\tau^{-1}_{\rm cas}=v_l/l_\bot$ and the ion-neutral collisional damping rate $|\omega_I|=\frac{\xi_n v_A^2k^2_\parallel}{2\nu_{ni}}$ in the strong MHD turbulence regime, is given by \citep{2015ApJ...810...44X,2016ApJ...826..166X}:
\begin{equation}
\label{eq.kc}
k_{\rm dam,\parallel}=
\frac{2\nu_{\rm ni}}{\xi_n}v_A^{-1},\\
\end{equation}
where $\xi_n=\rho_n/(\rho_i+\rho_n)$ is the fraction of neutrals. It holds for both sub-Alfv\'enic and super-Alfv\'enic turbulence. The perpendicular damping wavenumber $k_{\rm dam,\bot}$ can be derived from the anisotropy scaling in the strong turbulence regime:
\begin{equation}
\label{eq.kc_bot}
k_{\rm dam,\bot}=
\begin{cases}
(\frac{2\nu_{\rm ni}}{\xi_n})^{3/2}L_{\rm inj}^{1/2}v_{\rm inj}^{-3/2},~~~M_A>1\\
(\frac{2\nu_{\rm ni}}{\xi_n})^{3/2}L_{\rm inj}^{1/2}v_{\rm inj}^{-2}v_A^{1/2},~~~M_A<1\\
\end{cases}.
\end{equation}
It is important to note that the $k_{\rm dam,\bot}$ is the most crucial in determining the damping of the MHD turbulent cascade because the cascade mainly happens in the direction perpendicular to the local magnetic field. In addition, $k_{\rm dam,\bot}$ is larger than $k_{\rm dec,\parallel}$ (see Eq.~\ref{eq:decouple}), as damping of MHD turbulence takes place after neutrals decouple from ions. 

\begin{figure*}
\includegraphics[width=0.9\linewidth]{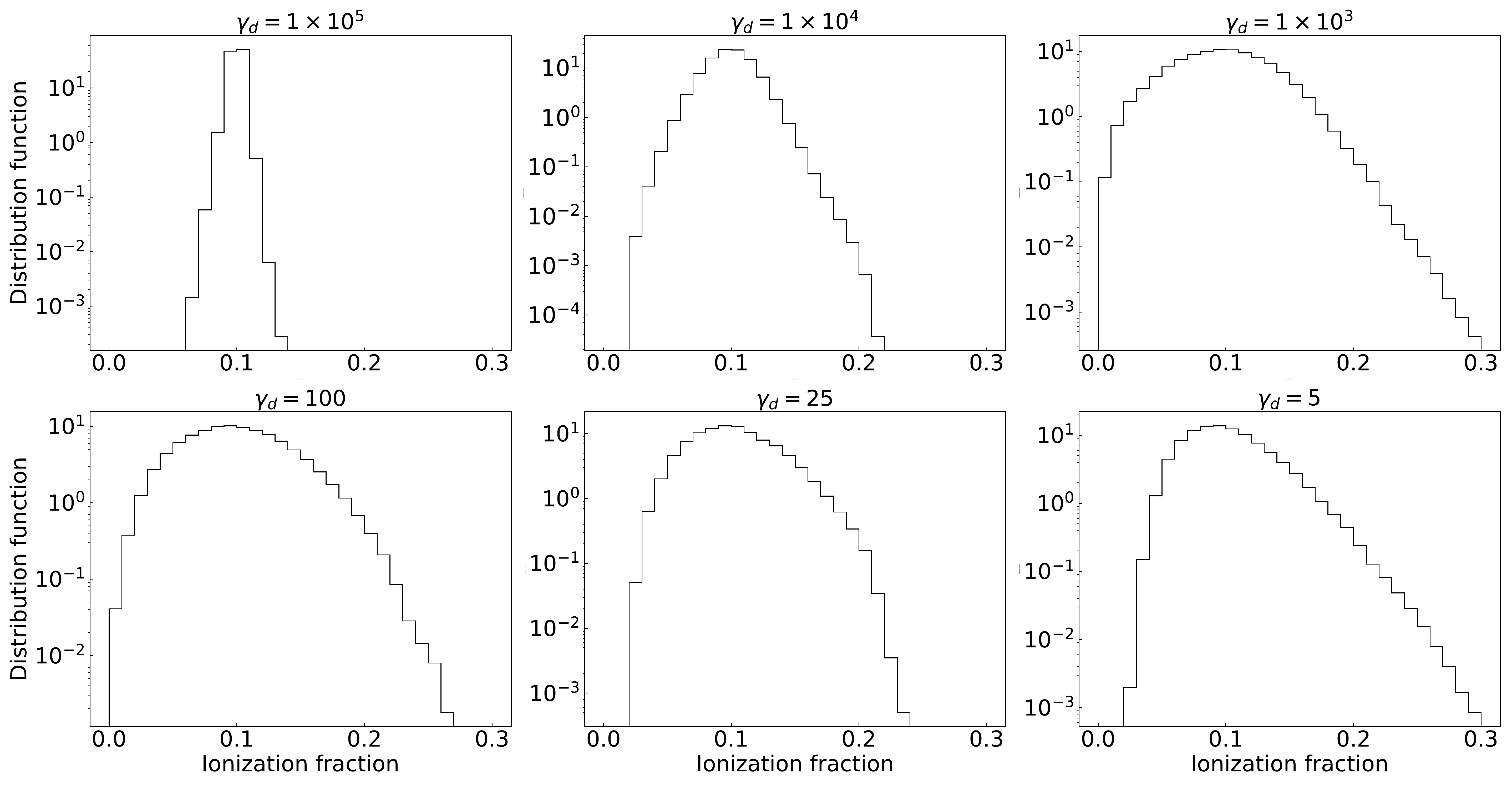}
    \caption{Normalized histogram of ionization fraction $\xi_i=\rho_i/(\rho_i+\rho_n)$. The fraction is calculated over the full simulation cube.}
    \label{fig:if_hist}
\end{figure*}

\begin{figure*}
\includegraphics[width=0.9\linewidth]{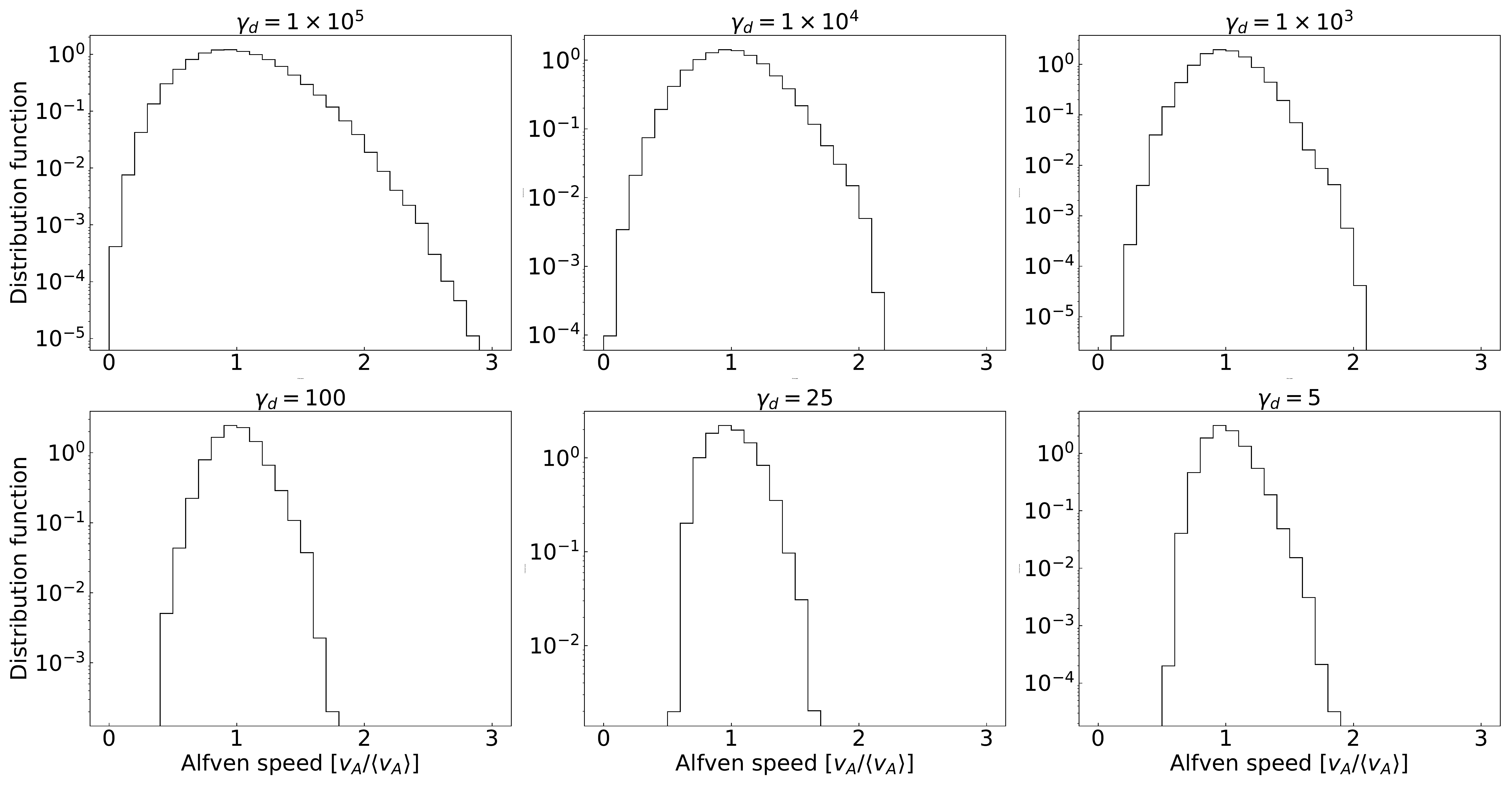}
    \caption{Normalized histogram of the local Alfv\'en speed $v_A=B/\sqrt{4\pi(\rho_i+\rho_n)}$ normalized by its mean value.}
    \label{fig:af_hist}
\end{figure*}

\section{Numerical results}
\label{sec:result}
\subsection{Velocity statistics}
\subsubsection{Neutral-ion collisional damping of MHD turbulence}
\textbf{Strongly coupled regime:} We present 2D slices of velocity fields for ions and neutrals in Figs.~\ref{fig:v_map1} and \ref{fig:v_map2}, taken perpendicular to the mean magnetic field at $x=240$ cell. In Fig.~\ref{fig:v_map1}, we show the cases of $\gamma_{\rm d}=1\times10^5$, $1\times10^4$, and $1\times10^3$, where $k_{\rm dec, \parallel}$ and $k_{\rm dec, \bot}$ are expected to be larger than the numerical dissipation wavenumber $k_{\rm dis}\approx40-50$, indicating that ions and neutrals are well-coupled over all length scales resolved in our simulations. Note $\gamma_{\rm d}$ values are given in numerical units. To obtain a dimensionless value, one can divide $\gamma_{\rm d}$ by $v_{\rm inj}/(L_{\rm inj}\rho_i)$, which is approximately 10 for the simulations with $\xi_i=0.1$. To calculate the theoretically expected $k_{\rm dec, \parallel}$ and $k_{\rm dec, \bot}$, as well as $k_{\rm dam, \parallel}$ and $k_{\rm dam, \bot}$, we adopt the mean values of density, magnetic field, and $v_{\rm inj}$ at $k=1$. We denote the $k_{\rm dec, \parallel}$ and $k_{\rm dec, \bot}$ as averaged decoupling wavenumbers. If the averaged decoupling wavenumbers are larger than the numerical dissipation wavenumbers, then neutrals and ions are on average well-coupled.

We find that the ion and neutral velocity structures are highly similar and exhibit anisotropy along the local magnetic field, with a stronger anisotropy toward a larger $k$ (see Fig.~\ref{fig:k_map_1e3}), similar to the anisotropy of MHD turbulence seen in a single fluid \citep{2000ApJ...539..273C,2002ApJ...564..291C,2019ApJ...878..157X}. The ion and neutral velocity spectra for the cases of $\gamma_{\rm d}=1\times10^5$ and $1\times10^4$ follow approximately the Kolmogorov scaling with a spectral slope of $-5/3$, while for $\gamma_{\rm d}=1\times10^3$, the spectra become a bit steeper, with a slope of $\approx-1.9$. Appendix Fig.~\ref{fig:k_map_1e3} further decomposes the velocity fields within different $k$ ranges and demonstrates the similarity in the velocity structures of ions and neutrals, irrespective of the length range.

\textbf{Transition from strongly to weakly coupled regime:} Fig.~\ref{fig:v_map2} presents the velocity distribution slices and turbulent kinetic energy spectra for three other setups with $\gamma_{\rm d}=100$ and $25$. We find that the spectra of ions and neutrals are different, and the spectrum of ions with a slope of approximately $-3.2$ is steeper than that of neutrals, indicative of more severe damping of the turbulent cascade in ions. The spectrum of neutrals is also steeper than the Kolmogorov one. It suggests that the neutral-ion decoupling does not happen sharply at a particular scale, but gradually over a range of scales. Compared to the strongly-coupled case, the velocity distributions of both neutrals and ions show a clear deficiency of small-scale structures, and the anisotropy is less apparent. We can also see that the neutral-ion decoupling does not happen at the ambipolar diffusion wavenumber $k_{\rm dec,\parallel}$. Instead, only at $k_{\rm dec,\bot}$, the spectra of ions and neutrals start to diverge. In Fig.~\ref{fig:k_map_25}, we decompose the ion and neutral velocity fields within different ranges of $k$ for the case with $\gamma_{\rm d}=25$ and find that differences in velocity structures appear starting from large $k$ values.

We note that previous studies with a low ionization fraction of $1\times10^{-4}$ suggest a Kolmogorov spectrum of neutrals no matter whether they are coupled or decoupled from ions \citep{2014MNRAS.439.2197M}. In our case, we have a higher ionization fraction of 0.1 and lower $\gamma_{\rm d}$. Our result suggests that the reduced $\gamma_{\rm d}$ may cause enhanced frictional damping and thus steepening of the spectra of neutrals and ions when they are coupled at $k<k_{\rm dec,\bot}$. At $k>k_{\rm dec,\bot}$, with relatively higher ion inertia neutrals are not fully decoupled from ions. Consequently, neutrals cannot develop a completely independent hydrodynamic cascade and their spectrum remains steep. This is, however, constrained by the limited internal range in our current numerical simulations. We expect the neutrals spectrum would become shallower at a sufficiently large wavenumber, in which neutrals are fully decoupled from ions. We note that our theoretically calculated decoupling and damping wavenumbers are based on the Kolmogorov scaling and scale-dependent anisotropy of Alfv\'enic turbulence. For strongly damped MHD turbulence with a steep spectrum and insignificant anisotropy, the theoretical estimates have a large uncertainty. Additional uncertainty comes from the fluctuations in the local ionization fraction in compressible MHD turbulence, and thus the decoupling of neutrals from ions does not happen on a single-length scale. We further discuss this point in \S~\ref{sssec:if}.

\textbf{Transition from weakly coupled to decoupled regime:} At $\gamma_{\rm d}$ = 5, neutrals are decoupled from ions on the turbulence injection scale, while ions are still globally coupled to neutrals up to $k\sim 5 - 10$. The velocity distributions of ions and neutrals exhibit differences, with neutrals displaying more isotropic velocity structures. In this regime, neutrals develop independent hydrodynamic turbulent cascades with a Kolmogorov slope, while ions undergo frequent collisions with neutrals, effectively damping the turbulence in ions. Therefore, ions exhibit a steep spectrum with a slope of approximately -2.6. The slope is related to the fraction of energy transferred to neutrals. Here we also see the ion spectrum exhibits a higher amplitude, most likely due to the artifact of driving turbulence. Our initial correlation timescale of the driving force is set to be equal to the crossing time of the Alfv\'en speed in the neutral and ion well-coupled cases, i.e. using the Alfv\'en speed calculated from the total density $\rho_n+\rho_i$. When neutrals decouple from ions, neutrals do not develop the Alfv\'en wave. The Alfv\'en speed then becomes larger due to smaller ion density. The correlation time fixed in the simulation is therefore too large for ions, so ions' turbulence cascades slower and gets higher velocity power. 

On the other hand, in Fig.~\ref{fig:drag}, we calculated the kinetic energy of ions and neutrals, the magnetic fluctuation energy, and the energy exchanged by their drag interaction. We found that the energy exchange in ions and neutrals are minimum in both $\gamma_{\rm d} = 10^5$ and $5$ cases. The velocity spectra are identical for the former, but the ion velocity spectrum has a higher amplitude for the latter. It suggests that the higher ion velocity power is not supplied by energy exchange with neutrals, but caused by the unequal correlation time of driving discussed above.


\begin{figure*}
\includegraphics[width=1.0\linewidth]{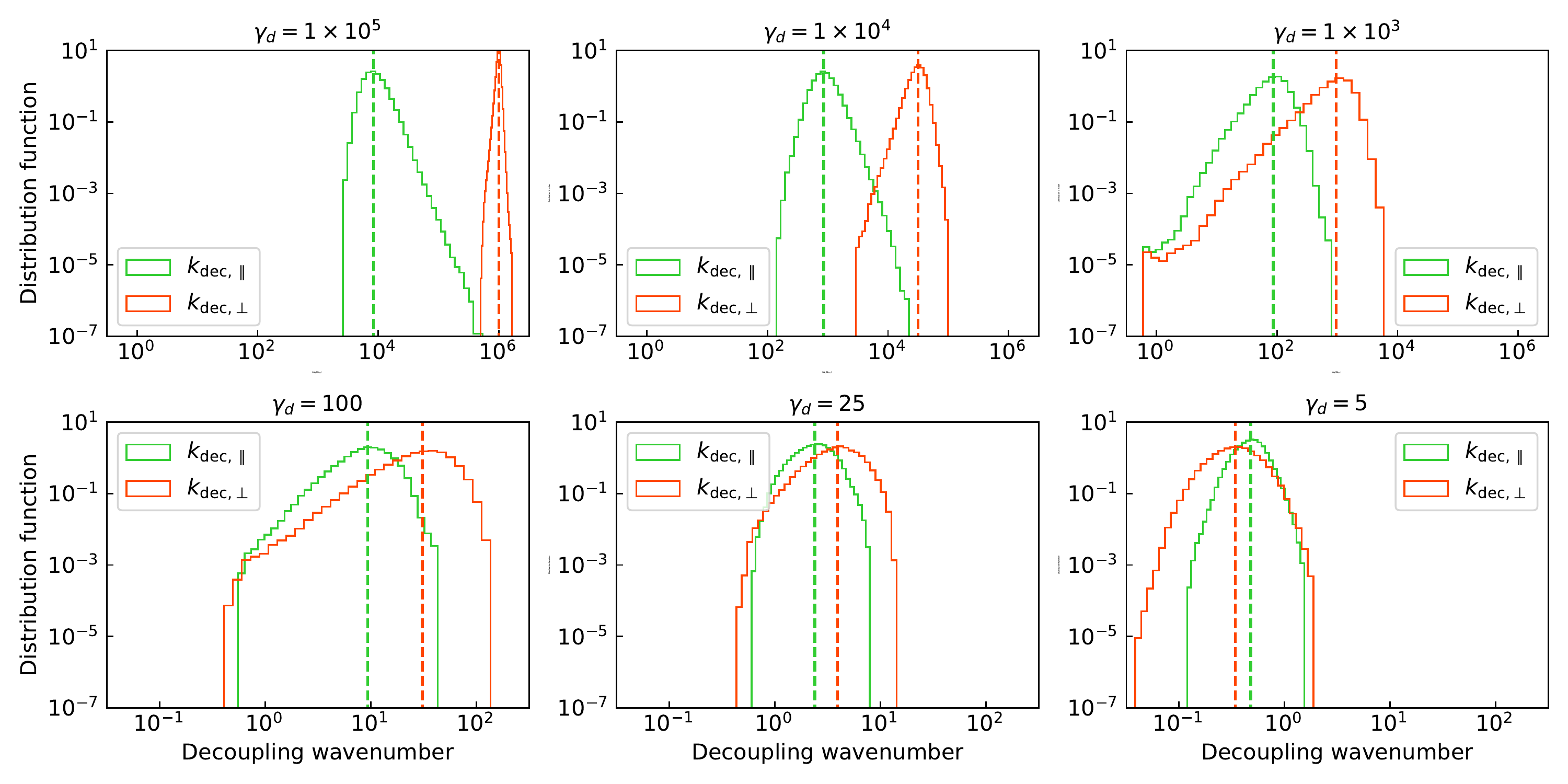}
    \caption{Normalized histogram of theoretically expected local neutral-ion parallel (green; $k_{\rm dec,\parallel}=\nu_{\rm ni}v_A^{-1}$) and perpendicular (orange; $k_{{\rm dec},\bot}=\nu_{\rm ni}^{3/2}L_{\rm inj}^{1/2}v_{\rm inj}^{-3/2}$) decoupling wavenumber in the unit of $2\pi/L_{\rm box}$. Dashed lines represent the median values.}
    \label{fig:k_hist}
\end{figure*}

\begin{figure*}
\includegraphics[width=0.95\linewidth]{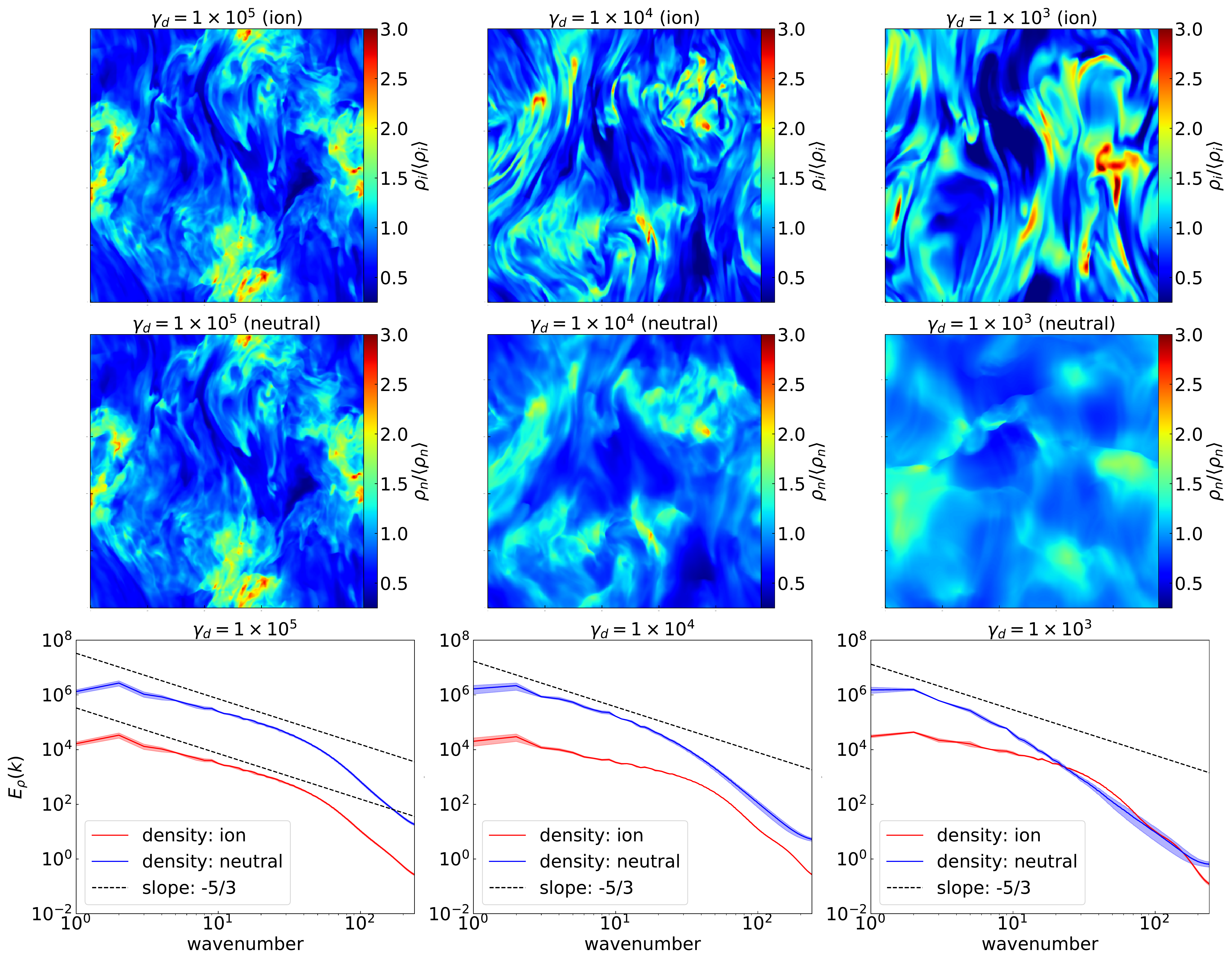}
    \caption{\textbf{Top and middle:} 2D slices (taken at $x=240$ cell) of ion (top) and neutral's (middle) density field. The density maps are normalized by the mean value. The view direction is perpendicular to the mean magnetic field, which is along the vertical direction. \textbf{Bottom:} Spectra of ion (red) and neutral's (blue) density. The spectra are averaged over several snapshots with one latest eddy turnover time. The shadowed areas represent the variations.}
    \label{fig:d_map1}
\end{figure*}

\begin{figure*}
\includegraphics[width=0.95\linewidth]{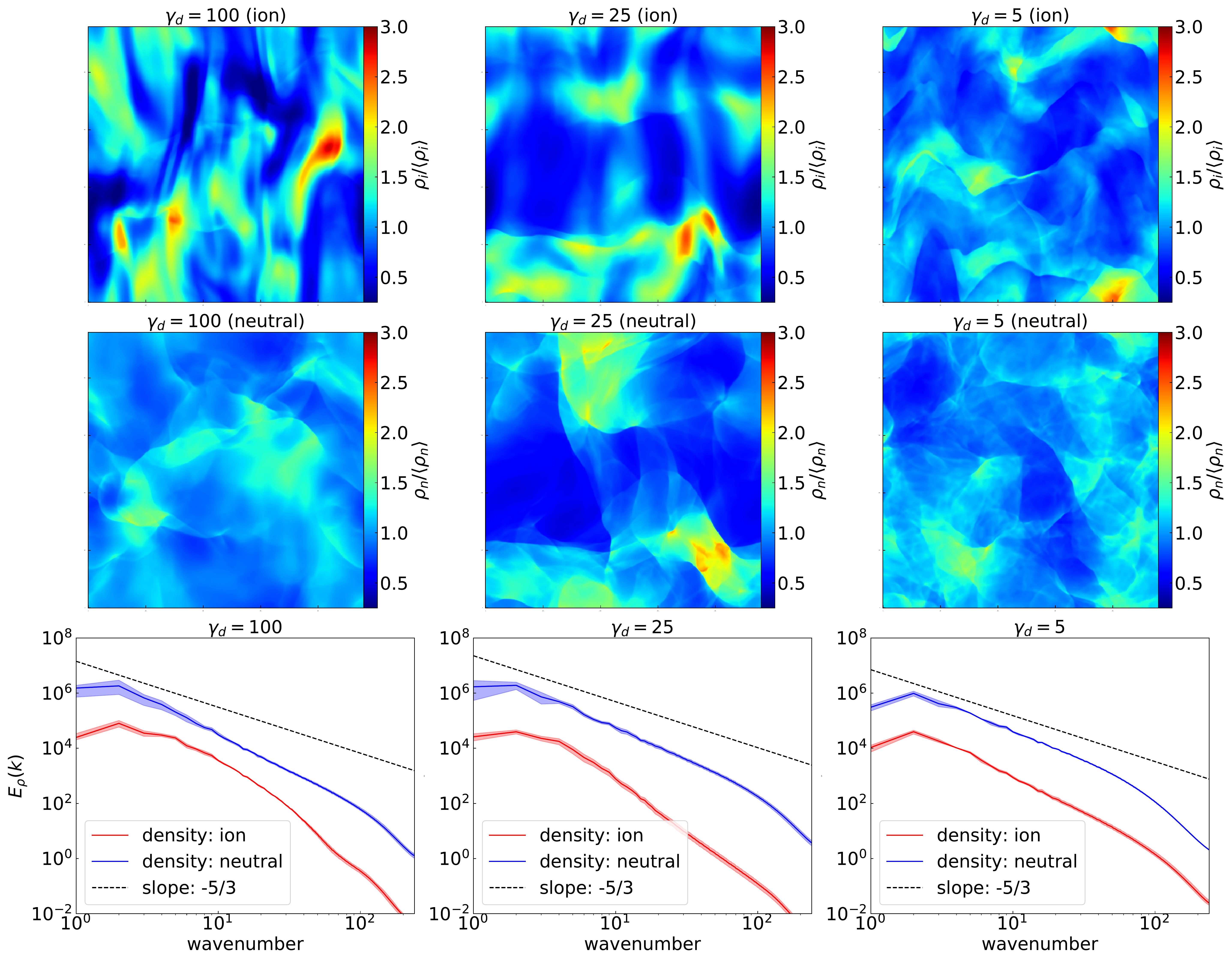}
    \caption{Same as Fig.~\ref{fig:d_map1}, but for $\gamma_{\rm d}=100,25,5$.}
    \label{fig:d_map2}
\end{figure*}
\subsubsection{Fluctuations in local ionization fraction, Alfv\'en speed, and decoupling scales}
\label{sssec:if}
The neutral-ion decoupling scale, as discussed in \S~\ref{sec:theory}, depends on the $\nu_{\rm ni} = \gamma_{\rm d} \rho_i= \gamma_{\rm d}\xi_i(\rho_i+\rho_n)$ (also $v_A$ for the parallel decoupling scale). Due to variations in density and magnetic fields in compressible MHD turbulence, these two quantities can exhibit significant fluctuations, resulting in local variations of the decoupling scales instead of a value.

To further investigate the variation of the local decoupling scale, we present histograms of the local ionization fraction in Fig.~\ref{fig:if_hist} with corresponding 2D slices shown in Fig.~\ref{fig:if_map}. The histogram of the $\gamma_{\rm d}=1\times10^5$ case is very narrow, with the ionization fraction concentrated around 0.1. However, as $\gamma_{\rm d}$ decreases, the ionization fraction starts to spread to both higher and lower values, indicating more significant local variations. For the other five cases with smaller $\gamma_{\rm d}$, we observe that the ionization fraction varies from approximately 0 to 0.3, while the global mean value of approximately 0.1 remains the same. These variations are due to fluctuations in ion and neutral densities. We expect that in supersonic turbulence with $M_s$ much larger than unity, where density fluctuations are more significant, the variation of ionization fraction may further increase.

In addition to the ionization fraction, we also investigate the local Alfv\'en speed fluctuations, shown in Fig.~\ref{fig:af_hist} with corresponding 2D slices in Fig.~\ref{fig:af_map}. The Alfv\'en speed fluctuations come from the variation of magnetic field strength and total density $\rho_i+\rho_n$. Unlike the ionization fraction, the case of $\gamma_{\rm d}=1\times10^5$ exhibits the widest histogram indicating significant variation of Alfv\'en speed. The histograms, however, become narrower for the other five cases with smaller $\gamma_{\rm d}$. Typically, we see the maximum value of $v_A/\langle v_A\rangle$ reaches $\sim2$ and minimum values are either $\sim0$ (for $\gamma_{\rm d}=1\times10^4$ and $1\times10^3$) or $\sim0.5$ (for $\gamma_{\rm d}=100,25,$ and $5$). 

The Alfv\'en speed and ionization fraction fluctuations result in local variations in the values of $k_{\rm dec,\parallel}$ and $k_{\rm dec,\bot}$. The distributions of their theoretically expected values calculated by using the local $\xi_i$ and $v_A$ are shown in Fig.~\ref{fig:k_hist}, which highlights the significant fluctuations that can occur. In the case of $\gamma_{\rm d}=1\times10^5$ and $1\times10^4$, the minimum values of $k_{\rm dec,\parallel}$ and $k_{\rm dec,\bot}$ are larger than the numerical dissipation wavenumber, suggesting that neutrals and ions remain locally well-coupled. Otherwise, if the local decoupling wavenumbers are smaller than the numerical dissipation wavenumber, neutrals are locally decoupled from ions. As seen in the case of $\gamma_{\rm d}=1\times10^3$, the local $k_{\rm dec,\parallel}$ can vary from $\approx1$ to $\approx1\times10^3$, indicating the existence of local decoupling. The local $k_{\rm dec,\bot}$ can even reach a larger value of $\approx8\times10^3$. Neutrals can fully decouple from ions only at wavenumbers larger than the maximum $k_{\rm dec,\bot}$.

For $\gamma_{\rm d}=100$, although the expected global mean decoupling scales are $k_{\rm dec,\parallel}\approx10$ and $k_{\rm dec,\bot}\approx32$, the local values of $k_{\rm dec,\parallel}$ and $k_{\rm dec,\bot}$ are also widely distributed from $\approx1$ to $\approx50$. When $\gamma_{\rm d}=25$, the range of $k_{\rm dec,\parallel}$ and $k_{\rm dec,\bot}$ is $\approx1$ to $\approx10$, and the damping of MHD turbulent cascade in neutrals due to local coupling with ions is more noticeable than the $\gamma_{\rm d}=100$ case. In this case, the neutral spectrum follows the steep spectrum of ions up to $k\sim10$, which is also the maximum value seen in the histogram of $k_{\rm dec,\bot}$. Finally, in the case of $\gamma_{\rm d}=5$, $k_{\rm dec,\parallel}$ and $k_{\rm dec,\bot}$ do not exceed 1.5 at their maximum, indicating that neutrals are fully decoupled from ions basically at all scales and develop a hydrodynamic turbulent cascade independently. 

These results suggest that neutral-ion decoupling does not occur on a single-length scale, but rather over an extended range of scales.

\subsection{Density statistics}
\label{ssec:density}
The local variation of the ionization fraction $\xi_i$ is important to understand the neutral-ion decoupling and damping of MHD turbulence. $\xi_i$ is directly related to the density fluctuations in ions and neutrals. In this section, we investigate the density statistics of ions and neutrals.

\begin{figure*}
\includegraphics[width=0.9\linewidth]{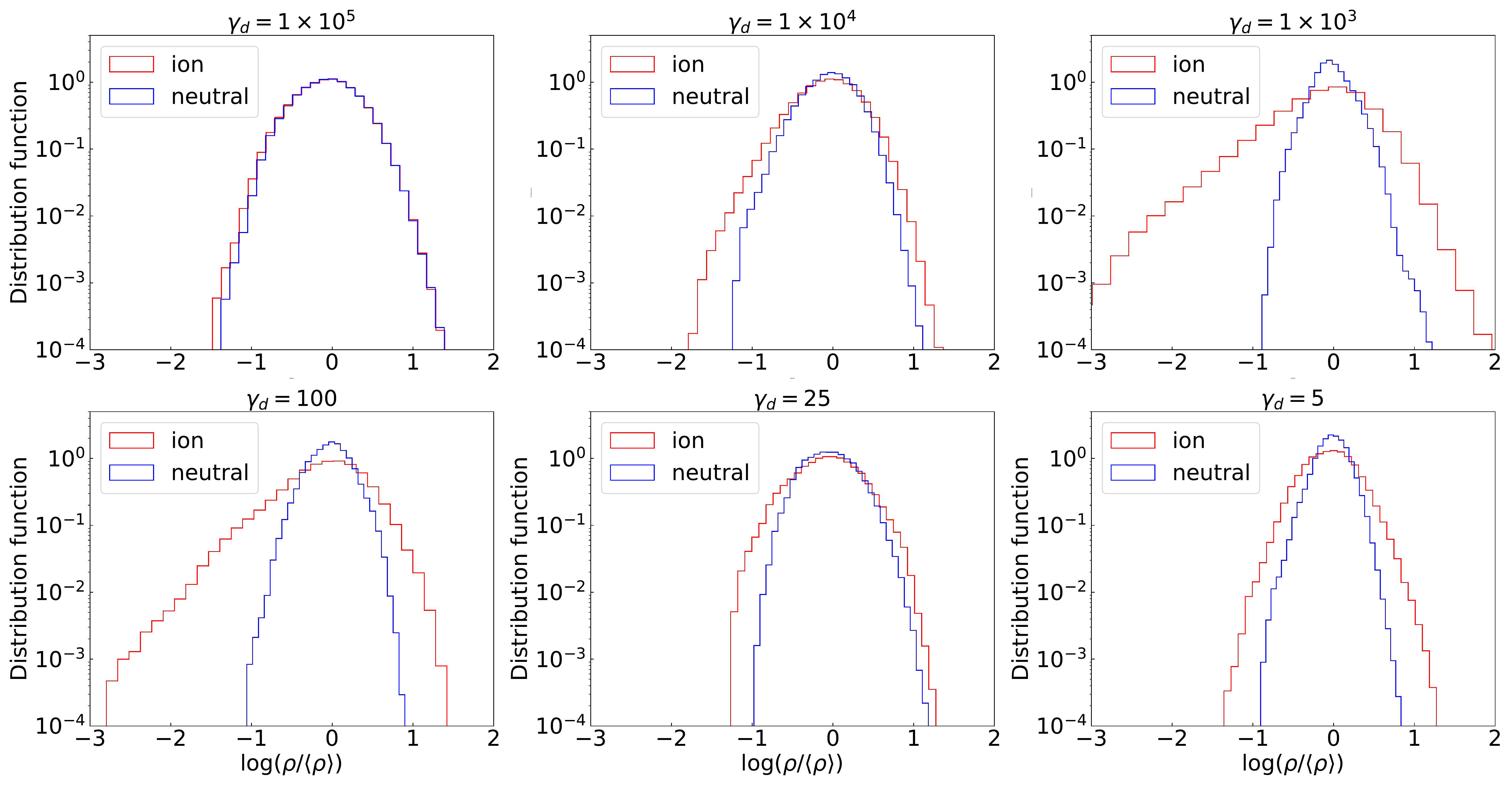}
    \caption{n-PDFs of ion (red) and neutral's (blue) logarithmic
    mass densities normalized by their mean densities.}
    \label{fig:pdf_rho}
\end{figure*}

\subsubsection{2D density distribution and density spectrum}
Figs.~\ref{fig:d_map1} and \ref{fig:d_map2} present 2D density slices (taken at $x=240$ cell, perpendicular to the mean magnetic field) and density spectra for ions and neutrals. When $\gamma_{\rm d}=1\times10^5$, we observe that the density distributions of ions and neutrals are nearly identical, with the structure regulated by turbulence anisotropy. Similar filamentary density structures are also seen in single-fluid MHD simulations \citep{2019ApJ...878..157X}. The spectra are a bit shallow, but they generally follow the Kolmogorov scaling, similar to their velocity spectra. However, when $\gamma_{\rm d}=1\times10^4$, the ion density distribution becomes different from that of neutrals. Ion density structures exhibit more apparent striations, while such small-scale structures are not seen in neutral density distribution. Correspondingly, the spectrum of the ion density becomes shallower (slope $\approx-1.1$), while that of the neutral density starts to become steeper at large $k$. These phenomena are more pronounced in the case of $\gamma_{\rm d}=1\times10^3$, where the slope of the ion density spectrum is $\approx-1.3$.

We see that despite the similar velocity structures seen in neutrals and ions in the strongly coupled regime, their density structures can differ significantly. The velocity field is likely to be dominated by incompressible Alfv\'enic turbulence, while density fluctuations are mainly induced by compressible turbulent motions. This can be seen from the difference in the velocity and density spectra of ions. Although neutrals are strongly coupled to the Alfv\'enic turbulent motions, they may be poorly coupled to the compressible MHD turbulent motions and thus do not exhibit the small-scale density structures created by the compressible MHD turbulent motions. 

Furthermore, when $\gamma_{\rm d}=100$ and $25$, the damping of MHD turbulence occurs (see Fig.~\ref{fig:v_map2}). The decoupling of neutrals from the Alfv\'enic turbulent motions also contributes to the difference in the density structures of neutrals and ions. The density distribution in neutrals appears isotropic. We observe that the anisotropic filamentary structures in ions become less apparent, which is due to the severe damping, while sharp-density jumps gradually appear on large scales. These sharp jumps are most significant in the neutral density. The spectra of both ions and neutrals are steep for $\gamma_{\rm d}=100$. Together with $\gamma_{\rm d}=1\times10^3$, these three cases are complicated because of the large variation of the local decoupling scale (see Fig.~\ref{fig:k_hist}). However, for $\gamma_{\rm d}=5$, the full decoupling of neutrals from ions is achieved, and only turbulence in ions is damped (see Fig.~\ref{fig:v_map2}). In this case, we clearly see that small-scale density structures arise in neutrals and the anisotropic filamentary structures in ions vanish, and the spectra are shallower (slope $\approx-2.6$ for ions and $\approx-2.1$ for neutrals) than those at $\gamma_{\rm d}=100$ (slope $\approx-3.4$ for ions and $\approx-2.7$ for neutrals) and $\gamma_{\rm d}=25$ (slope $\approx-3.9$ for ions and $\approx-2.3$ for neutrals). 

\subsubsection{The probability distribution function of the logarithmic mass density }
We present the probability distribution function (n-PDF) of the logarithmic mass densities of ions and neutrals, normalized by their respective mean values, as shown in Fig.~\ref{fig:pdf_rho}. The n-PDF is a widely used tool for studying density statistics in single-fluid MHD turbulence \citep{2011ApJ...727L..21P,2018ApJ...863..118B}, as it directly reveals the significance of density fluctuations. In general, the minimum and maximum values of the neutrals' n-PDF are approximately -1 and 1, respectively. These values vary a bit at large $\gamma_{\rm d}=1\times10^5$ and $1\times10^4$. As in the case of single-fluid turbulence, we expect the width of the n-PDF to be correlated with $M_s$ \citep{2002ApJ...576..870P}. A high sonic Mach number, especially greater than unity, is typically associated with shocks, which lead to high-density contrasts and a more dispersed n-PDF. This behavior is commonly observed in studies of single-fluid MHD turbulence \citep{2011ApJ...727L..21P,2018ApJ...863..118B}.

When $\gamma_{\rm d}=1\times10^5$, the ions' n-PDF closely resembles that of neutrals. However, the ions' n-PDFs become more dispersed with smaller $\gamma_{\rm d}$. While the maximum value of the ions' n-PDFs remains stable at ${\rm log(\rho/\langle\rho\rangle)}\approx1.25 - 2.0$, the minimum value reaches ${\rm log(\rho/\langle\rho\rangle)}\approx-3.0$ for $\gamma_{\rm d}=1\times10^3$ and 100. This suggests that the ion density exhibits significant local fluctuations. The ions' n-PDFs narrow again for $\gamma_{\rm d}=25$ and 5, with ${\rm log(\rho/\langle\rho\rangle)}\approx-1.25$ at the minimum.

\subsection{Magnetic field statistics}
\begin{figure*}
\includegraphics[width=1.0\linewidth]{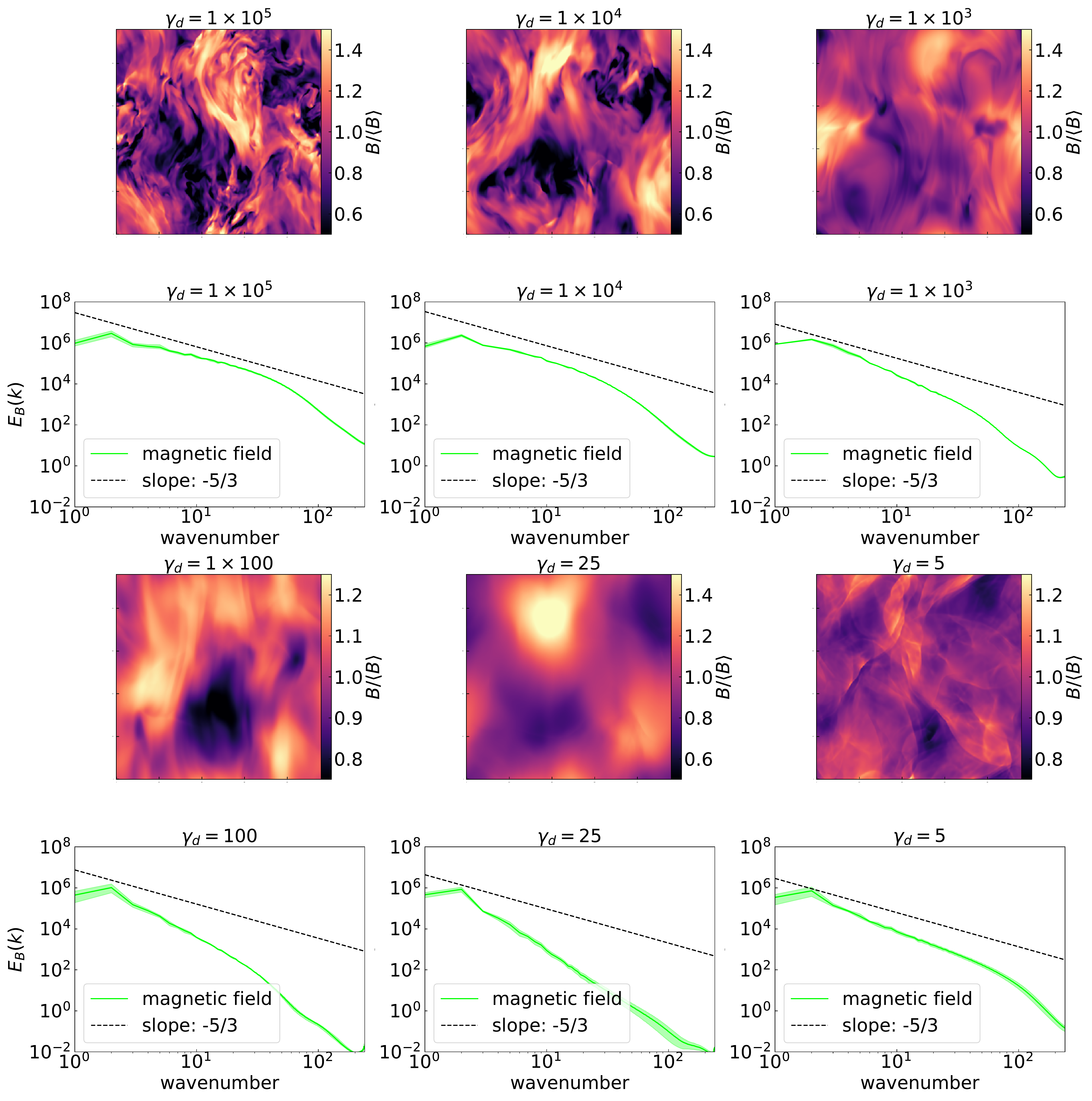}
    \caption{\textbf{$1^{\rm st}$ and $3^{\rm rd}$:} 2D slices (taken at $x=240$ cell) of magnetic field strength. The magnetic field maps are normalized by the mean value. The view direction is perpendicular to the mean magnetic field, which is along the vertical direction. \textbf{$2^{\rm nd}$ and $4^{\rm th}$:} magnetic energy spectrum. The spectra are averaged over several snapshots within one eddy turnover time. The shadowed areas represent the variations.}
    \label{fig:b_map}
\end{figure*}

Fig.~\ref{fig:b_map} displays 2D slices of total magnetic field strength taken at $x=240$ cell perpendicular to the mean magnetic field direction, as well as magnetic energy spectra calculated for the full cube. In the neutral-ion locally well-coupled state, where $\gamma_{\rm d}=1\times10^5$ and $1\times10^4$, the magnetic field fluctuations elongate anisotropically along the magnetic field direction, akin to the velocity and density structures shown in Figs.\ref{fig:v_map1} and ~\ref{fig:d_map1}. The spectra exhibit Kolmogorov scaling overall. However, when $\gamma_{\rm d}=1\times10^3$, the magnetic field structures are less filamentary, and the spectrum becomes steeper with a slope of $\approx-2.8$ than that of the velocity spectra. This steepening of the magnetic energy spectrum when the velocity spectra of neutrals and ions are similar has been observed also in \cite{2014MNRAS.439.2197M}. It may be attributed to the effect of local neutral-ion decoupling. Alternatively, the fast modes in MHD turbulence may get damped at $k$ smaller than the damping scale of Aflv\'en modes \citep{2015ApJ...810...44X,2016ApJ...826..166X,2018arXiv181008205X}. The damping of the magnetic fluctuations generated by fast modes may result in a steeper magnetic energy spectrum than the kinetic energy spectrum that is dominated by the Alfv\'en modes.

In the weakly coupled regime with $\gamma_{\rm d}=100$ and $25$, small-scale magnetic field structures are less prominent, and the spectra become even steeper, indicating that the magnetic field energy becomes concentrated on larger scales. This is naturally expected due to the severe neutral-ion collisional damping. However, when the neutral-ion decoupling occurs at the injection scale (i.e., $\gamma_{\rm d}=5$), the situation changes. The spectrum becomes shallower compared to the cases of $\gamma_{\rm d}=1\times10^3$, $100$, and $25$ (slope $\approx-2.8, -3.6$, and $-4.0$,  respectively). The slope is close to $-2.23$. It suggests a weak damping effect, as seen in Fig.~\ref{fig:v_map2}. Overall, we see that the magnetic energy spectrum has a similar shape as the turbulent kinetic energy spectrum in ions.

\section{Discussion}
\label{sec:dis}
\subsection{Comparison with earlier studies}
The two-fluid (neutral and ion) simulation requires a very short time step to stably accommodate the fastest wave speed in the problem \citep{2014MNRAS.439.2197M}. This is computationally expensive since the low ionization fraction in the ISM (typically $10^{-4}-10^{-7}$ for cold molecular clouds, see \citealt{2005pcim.book.....T,2011piim.book.....D}) results in an extremely large Alfv\'en speed. The practical time step is even smaller because the Alfv\'en speed in a simulation has its own variations (see Fig.~\ref{fig:af_map}). The stable time step is therefore determined by the largest Alfv\'en speed. Consequently, this computational challenge limits the numerical study of two-fluid MHD turbulence. In the study of two-fluid MHD turbulence, some earlier research has focused on the damping of MHD turbulence using simulations generated by the RIEMANN code \citep{2010MNRAS.406.1201T,2014MNRAS.439.2197M,2015ApJ...805..118B}. These studies have primarily concentrated on supersonic ($M_s>1$) 

In our study, we used the Athenak code to conduct trans-sonic ($M_s\approx1$) two-fluid simulations with a moderately low ionization fraction of 0.1. 
To further examine the effect of ionization fraction, a comparison with a lower ionization fraction of 0.01 is presented in Fig.~\ref{fig:if001}. Compared to earlier studies using a fixed $\gamma_{\rm d}$, this approach of using reduced and varying $\gamma_{\rm d}$ enables us to study MHD turbulence in different coupling regimes with far fewer computational resources. 
Compared to earlier studies, we report newly discovered properties of two-fluid MHD turbulence. We quantitatively compared our numerical measurements with previous theoretical predictions on neutral-ion decoupling and damping scales \citep{2015ApJ...810...44X,2016ApJ...826..166X}, and found their large variations due to the large density fluctuations of ions and neutrals. The computational tools' application to other physical environments, like the very local ISM, will be explored in our future work.

\subsection{Implications for related studies}
\subsubsection{Cosmic ray transport}
The damping of MHD turbulence is crucial for understanding the transport of cosmic rays (CRs) in the multi-phase ISM. The damping of MHD turbulence should be taken into account for both resonant scattering \citep{2013ApJ...779..140X,HLX21a} and non-resonant mirroring \citep{2021ApJ...923...53L}. The severe damping of MHD turbulence in a weakly ionized medium can significantly affect the efficiency of scattering and the spatial confinement of CRs \citep{2016ApJ...826..166X}. 

\subsubsection{Velocity gradient}
Determining the scale at which neutrals and ions become decoupled is a challenging task in observations. In particular, it is non-trivial to obtain the velocity spectra of ions and neutral. However, this can be achieved by the Velocity Gradient Technique (VGT; \citealt{LY18a,HYL18}), which is a new approach to tracking magnetic fields using spectroscopic data. VGT is based on the anisotropy of MHD turbulence, where turbulent eddies align themselves along the magnetic fields. Velocity gradient serves as a detector of the anisotropy and, therefore, can reveal the magnetic field direction. 

The study shows that this anisotropy is absent when neutrals and ions become decoupled. We can expect that at a length scale larger than the decoupling scale, neutral turbulence and ion turbulence act as a single fluid and exhibit anisotropy, with velocity gradients of both species oriented in the same direction. At smaller length scales where neutrals decouple from ions, their velocity fields change (see Fig.~\ref{fig:v_map2}) so that the relative orientation of their velocity gradients changes. Therefore, comparing the directions of the (ions and neutrals) velocity gradients at different length scales can independently reveal the neutral-ion decoupling scale (i.e., perpendicular ambipolar diffusion scale). This approach could provide unique constraints on the important perpendicular ambipolar diffusion scale in observation for a better understanding of star formation \citep{1956MNRAS.116..503M,1972ApJ...173...87N}.

\section{Summary}
\label{sec:con}
Magnetized turbulence is ubiquitous in the partially ionized ISM. The interaction between neutral and ionized species can modify the properties of MHD turbulence and cause neutral-ion collisional damping. On the basis of the two-fluid MHD turbulence simulations generated from the AthenaK code, we numerically studied the statistical properties of velocity, density, and magnetic field in different regimes of ion-neutral coupling. Our main findings are:
\begin{enumerate}
    \item Our results demonstrate that in the (neutral-ion) strongly coupled regime, velocity statistics in the two-fluid simulations can resemble those in single-fluid MHD turbulence.
    \item In the weakly coupled regime, we observe that damping of turbulence can occur in both, resulting in their steep kinetic energy spectra compared to the Kolmogorov spectrum, while the damping of turbulence in ions is more severe. We find that due to the large density fluctuations in ions and neutrals, the ionization fraction has a large spatial variation, which causes a significant local variation of the neutral-ion decoupling scale. As a result, both neutral-ion decoupling and damping of MHD turbulence happen over a range of length scales.
    \item In the transition regime from weakly coupled to decoupled regime ($\gamma_{\rm d}=5$), the damping of MHD turbulence takes place in ions showing a steep kinetic energy spectrum. Neutrals develop an independent hydrodynamic turbulent cascade and the corresponding kinetic energy follows the Kolmogorov scaling.
    \item We find that in the strongly coupled regime with similar velocity structures in ions and neutrals, their density structures can exhibit significant differences. The small-scale enhanced density fluctuations seen in ions are absent in neutrals, and the density spectrum of ions is also much shallower than that of neutrals. This may be caused by the poor coupling of neutrals to the compressive MHD turbulent motions. 
    \item We show that the probability distribution function (n-PDF) of neutral mass density is insensitive to the coupling status between ions and neutrals. The n-PDF of the ions is broader than that of the neutrals, and its width varies in different coupling regimes.
    \item Finally, we find that the magnetic energy spectrum in general has a similar shape as the kinetic energy spectrum of ions. 
\end{enumerate}

\section*{Acknowledgements}
Y.H. and A.L. acknowledge the support of NASA ATP AAH7546, NSF grants AST 2307840, and ALMA SOSPADA-016. Financial support for this work was provided by NASA through award 09\_0231 issued by the Universities Space Research Association, Inc. (USRA). S.X. acknowledges the support for this work provided by NASA through the NASA Hubble Fellowship grant \# HST-HF2-51473.001-A awarded by the Space Telescope Science Institute, which is operated by the Association of the Universities for Research in Astronomy, Incorporated, under NASA contract NAS5-26555. This work used SDSC Expanse CPU in SDSC through the allocation PHY230032 from the Advanced Cyberinfrastructure Coordination Ecosystem: Services \& Support (ACCESS) program, which is supported by National Science Foundation grants \#2138259, \#2138286, \#2138307, \#2137603, and \#2138296. DA acknowledges the Nazarbayev University Faculty Development Competitive Research Grant Program \#11022021FD2912.

\section*{Data Availability}
The data underlying this article will be shared on reasonable request to the corresponding author.


\bibliographystyle{mnras}
\bibliography{example} 
\appendix
\section{Velocity maps at different $k$ ranges}
\begin{figure*}
\includegraphics[width=1.0\linewidth]{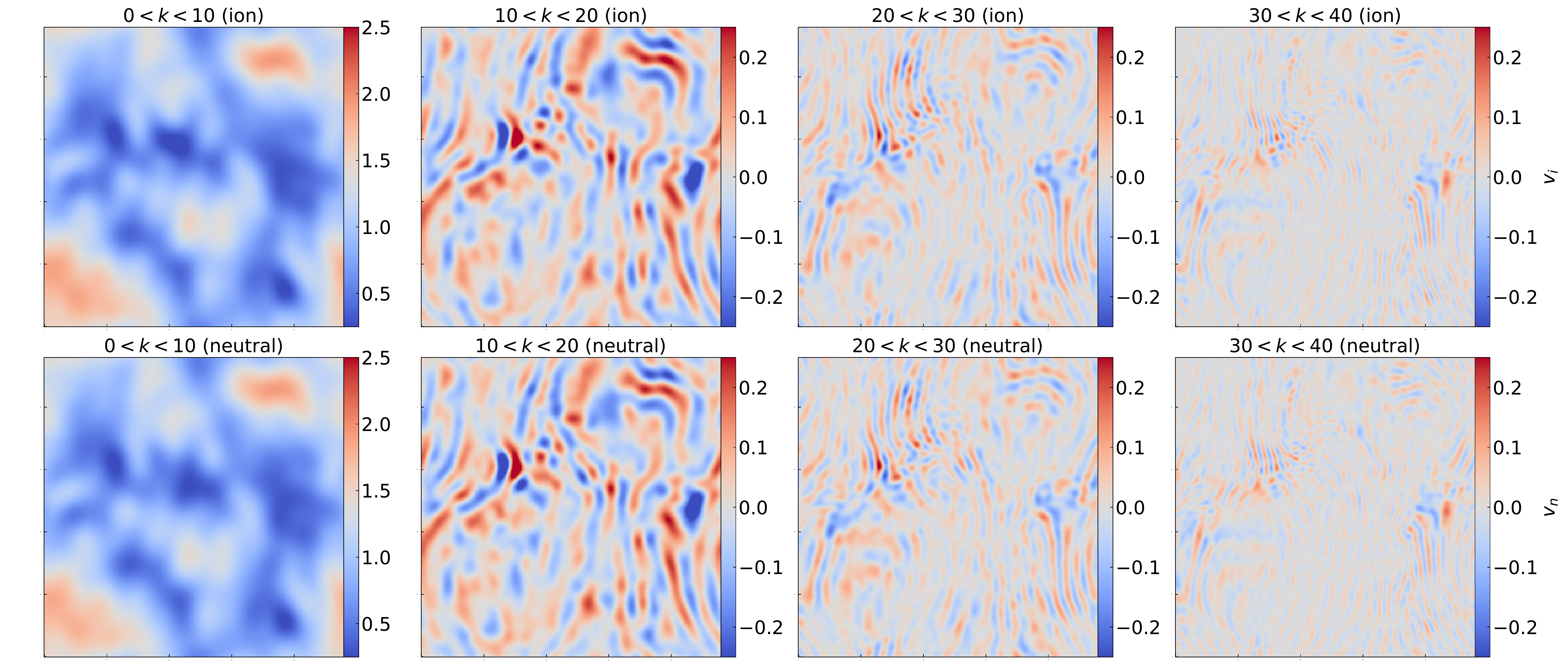}
    \caption{2D slices (taken at $x=240$ cell) of ion velocity (top) and neutral velocity (bottom) for the simulation of $\gamma_{\rm d}=1\times10^3$. The velocity fields are decomposed into different $k$ ranges in Fourier space and then transformed back to real space.}
    \label{fig:k_map_1e3}
\end{figure*}

\begin{figure*}
\includegraphics[width=1.0\linewidth]{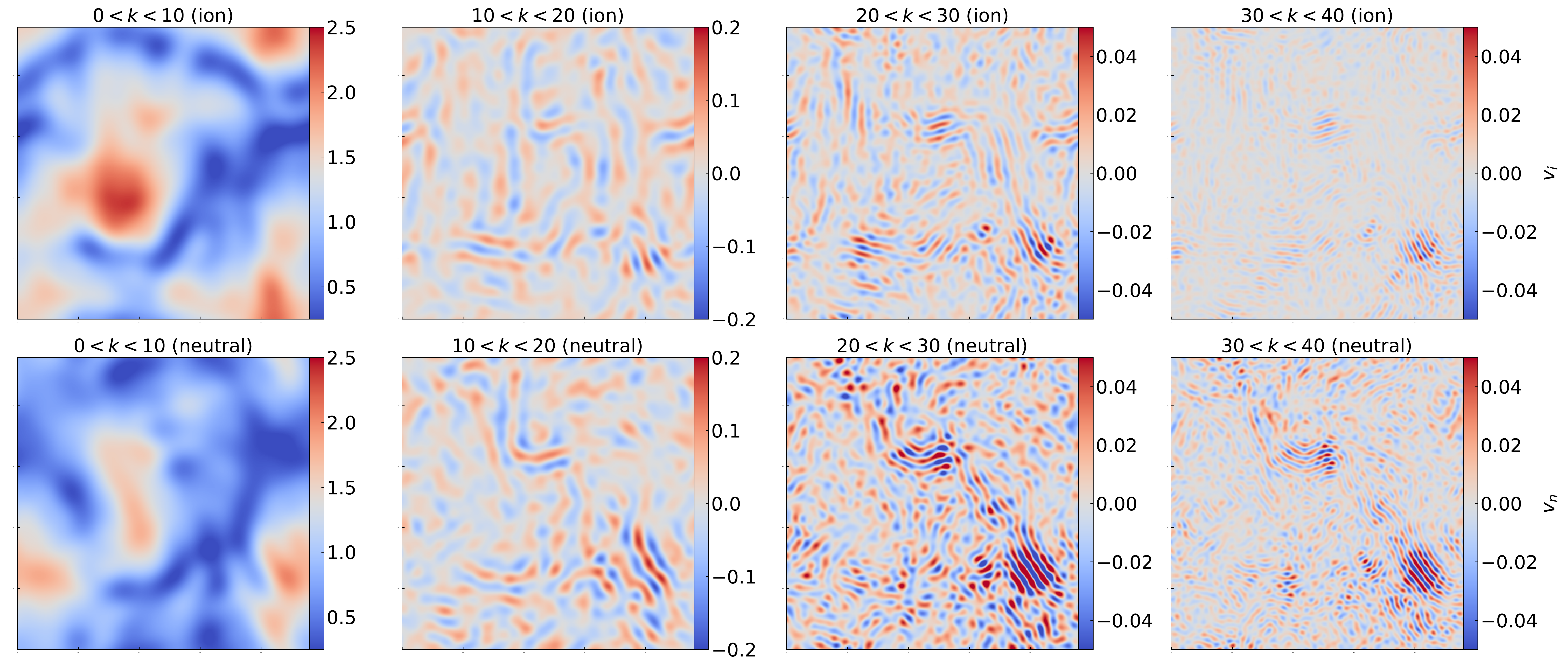}
    \caption{Same as Fig.~\ref{fig:k_map_25}, but for $\gamma_{\rm d}=25$.}
    \label{fig:k_map_25}
\end{figure*}
Figs.~\ref{fig:k_map_1e3} and \ref{fig:k_map_25} offer insights into the velocity structure of ions and neutrals for different $k$ ranges at $\gamma_{\rm d}=1\times10^3$ and $25$, respectively. To produce these results, we applied a Fourier transformation to the velocity cube and excluded velocity values outside the $k$ ranges of interest before retransforming the processed cubes back to real space.

For $\gamma_{\rm d}=1\times10^3$, the velocity slices of the ions and neutrals are nearly identical across all ranges $k$. Anisotropic velocity structures become more prominent at higher $k$ values. While ions and neutrals may exhibit local decoupling, they are globally well-coupled. On the contrary, at $\gamma_{\rm d}=25$, a distinct difference is observed between ions and neutrals, starting from small $0<k<10$. At this value of $\gamma_{\rm d}$, ions and neutrals are globally decoupled from each other

\section{Comparison with lower ionization fraction}
\begin{figure*}
\includegraphics[width=1.0\linewidth]{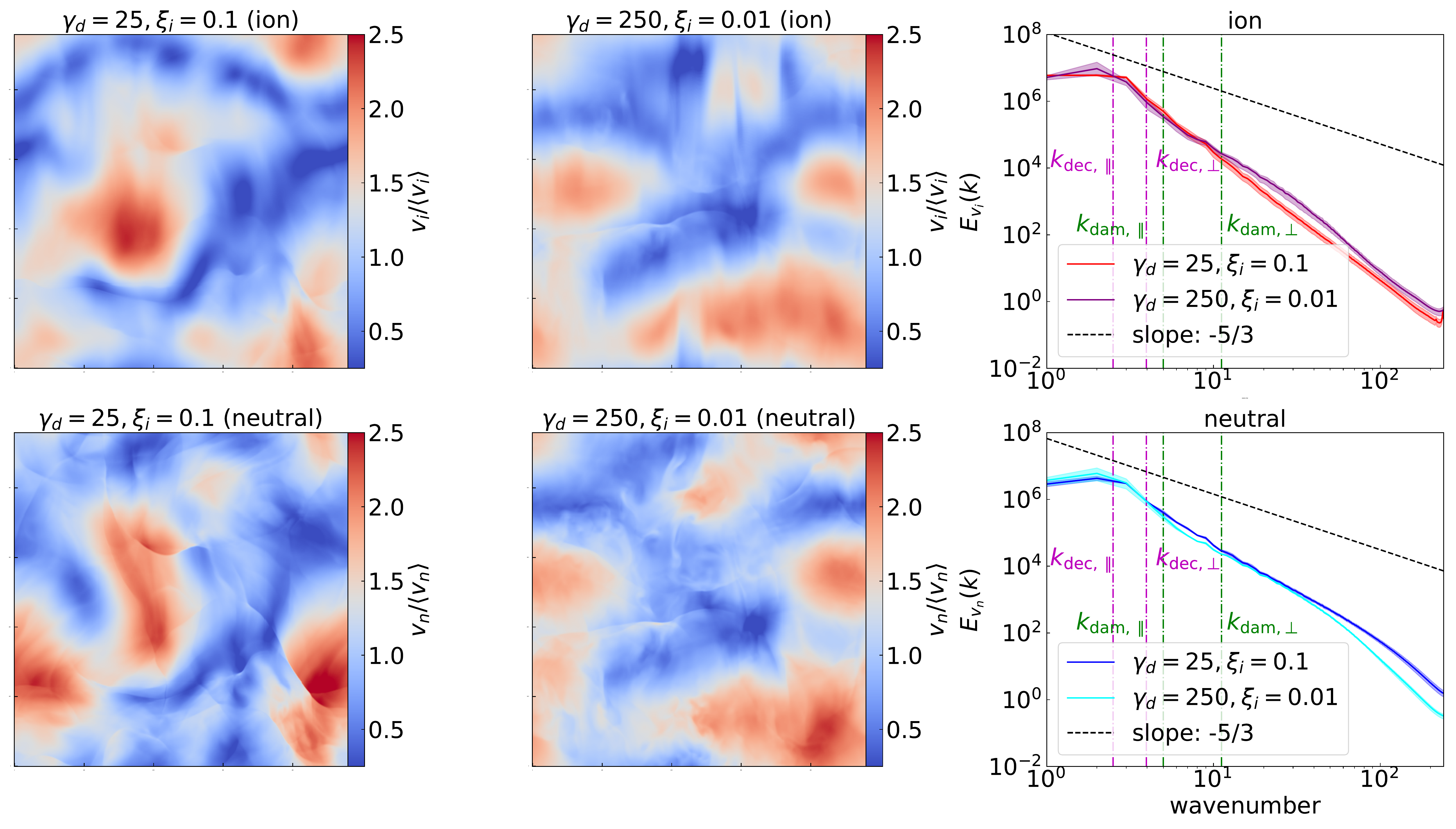}
    \caption{\textbf{Left and middle:} 2D slices (taken at $x=240$ cell) of ion (top) and neutral's (middle) velocity field with different drag coefficient $\gamma_{\rm d}$ and ionization fraction $\xi_i$. Velocity maps are normalized by the mean value. The view direction is perpendicular to the magnetic field, which is along the vertical direction. \textbf{Right:} Kinetic energy spectra of ion (top) and neutral (bottom). The shadowed areas represent the variations. Purple and green lines represent the expected neutral-ion decoupling and damping scales, respectively.}
    \label{fig:if001}
\end{figure*}

Fig.~\ref{fig:if001} presents the comparison of two $\gamma_{\rm d}$ and $\xi_i$ combinations. The values of $\gamma_{\rm d}=25, 250$ and $\xi_i=0.1, 0.01$ are meticulously selected so that the neutral-ion collisional frequency remains consistent. For the 2D velocity slices, the ion and neutral maps exhibit morphological differences. Furthermore, we calculate the velocity spectra, as shown in Fig.~\ref{fig:if001}. The kinetic energy spectrum of the ions (for $\gamma_{\rm d}=250$ and $\xi_i=0.01$) becomes slightly shallower when the wavenumber is larger than $k_{\rm dam,\bot}$. However, we do not observe any noticeable differences in the kinetic energy spectrum of neutrals at wavenumbers smaller than the dissipation. 

\section{Fluctuations of ionization fraction and Alfv\'en speed}
\begin{figure*}
\includegraphics[width=0.99\linewidth]{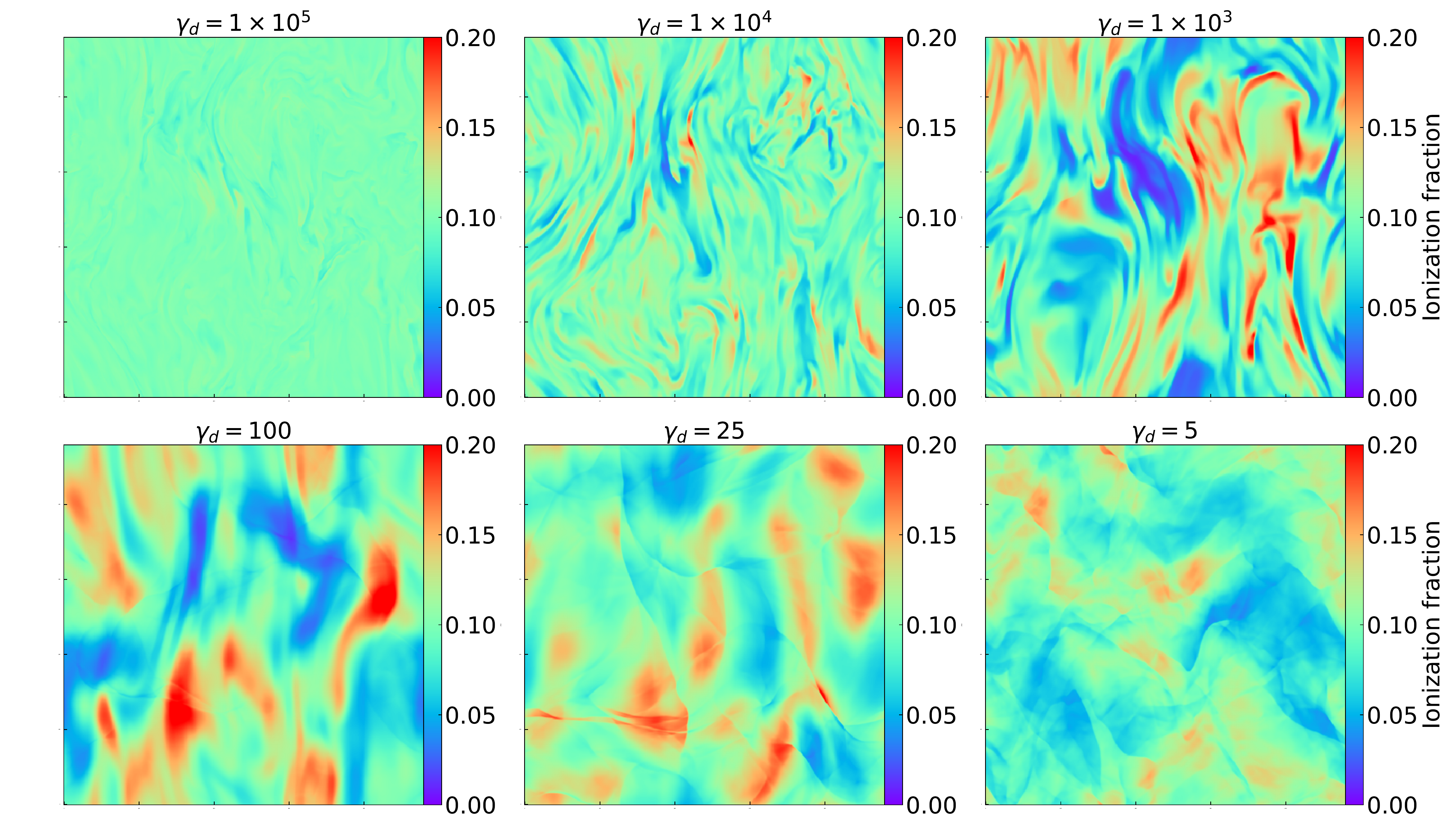}
    \caption{2D slices (taken at $x=240$ cell) of ionization fraction $\xi_i=\rho_i/(\rho_i+\rho_n)$.}
    \label{fig:if_map}
\end{figure*}

\begin{figure*}
\includegraphics[width=0.99\linewidth]{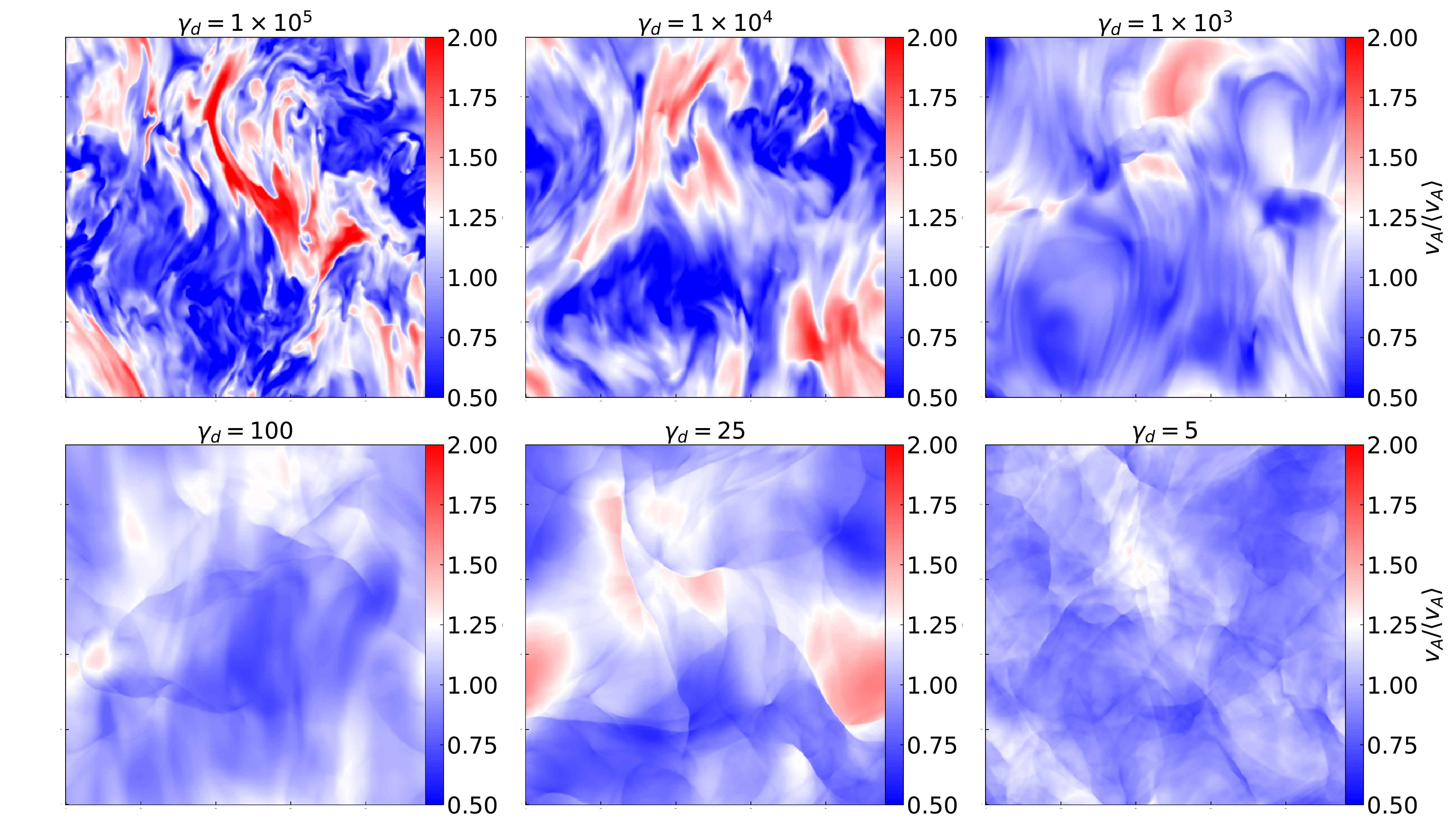}
    \caption{2D slices (taken at $x=240$ cell) of local Alfv\'en speed $v_A=B/\sqrt{4\pi(\rho_i+\rho_n)}$ normalized by its mean value.}
    \label{fig:af_map}
\end{figure*}
The ionization fraction $\xi_i=\rho_i/(\rho_i+\rho_n)$ and Alfv\'en speed $v_A$ are essential physical quantities for understanding neutral-ion decoupling. In Figs.~\ref{fig:if_map} and \ref{fig:af_map}, we present 2D distributions of these two parameters.

For $\xi_i$, we observe that its fluctuation is the smallest when $\gamma_{\rm d}=1\times10^5$. However, as $\gamma_{\rm d}$ decreases, the ionization fraction varies significantly from 0 to 0.3. On the contrary, $v_A$ has strong fluctuations for $\gamma_{\rm d}=1\times10^5$, but the fluctuations become weaker as $\gamma_{\rm d}$ decreases. We notice very apparent sharp $v_A$-jump edges in the cases of $\gamma_{\rm d}=100,25$, and 5. These edges are contributed by the jumps in density (see Fig.~\ref{fig:d_map2}).

\section{Energy variation with different $\gamma_{\rm d}$}
\begin{figure}
\includegraphics[width=0.99\linewidth]{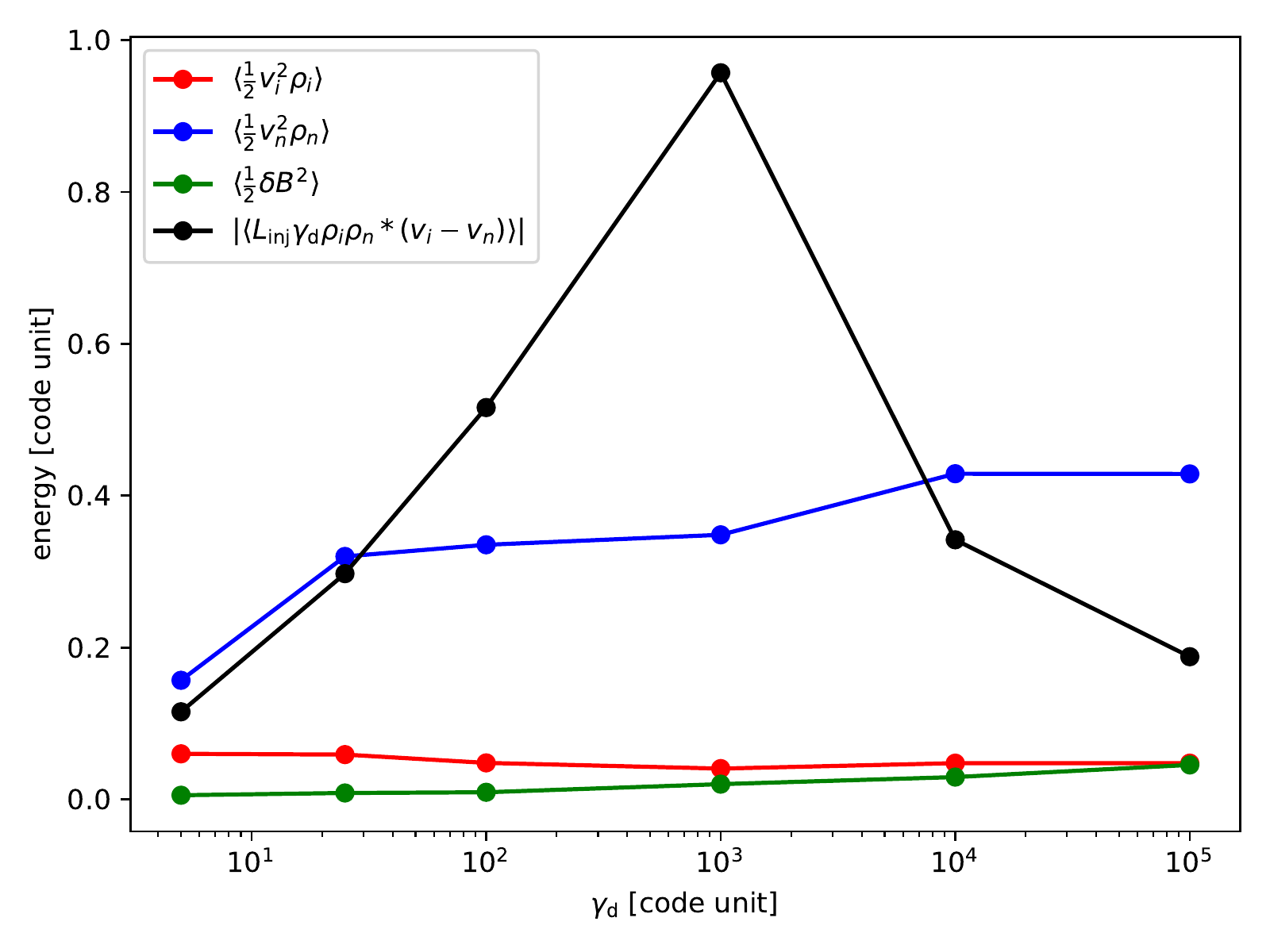}
    \caption{The variation of ions' and neutrals' kinetic energy, magnetic field fluctuation energy, and the energy exchanged by their drag interaction, as a function of $\gamma_{\rm d}$.}
    \label{fig:drag}
\end{figure}

Fig.~\ref{fig:drag} presents the ions' and neutrals' kinetic energy, magnetic field fluctuation energy, and the energy exchanged by their drag interaction, averaged over the simulation box. It shows neutrals always have higher kinetic energy due to their large density. The kinetic energy of the ions increases at small $\gamma_{\rm d}$, but the kinetic energy of the neutrals decreases. This is caused by the imbalance of the driving force's correlation time and Aflv\'en speed in the neutral-ion decoupled regime, see \S~\ref{sec:result}. On the other hand, the energy of magnetic field fluctuations cannot exceed the ions; kinetic energy, because the magnetic field fluctuation is induced by turbulent velocity.

We noticed that the energy change between ions and neutrals is minimal in the fully coupled (i.e., $\gamma_{\rm d}=10^5$) and fully decoupled (i.e., $\gamma_{\rm d}=5$) cases. It, however, achieves maximum when $\gamma_{\rm d}=10^3$, in which neutrals and ions start to decouple locally in terms of velocity.

\bsp	
\label{lastpage}
\end{document}